\documentclass[12pt]{article}
\usepackage{amssymb,amsmath,epsfig}
\allowdisplaybreaks

\begin{document}

\title{\bf Effect of Extended Gravitational Decoupling on Isotropization and Complexity in
$f(\mathbb{R},\mathbb{T})$ Theory}
\author{M. Sharif$^1$ \thanks{msharif.math@pu.edu.pk} and Tayyab Naseer$^{1,2}$ \thanks{tayyabnaseer48@yahoo.com}\\
$^1$ Department of Mathematics and Statistics, The University of Lahore,\\
1-KM Defence Road Lahore, Pakistan.\\
$^2$ Department of Mathematics, University of the Punjab,\\
Quaid-i-Azam Campus, Lahore-54590, Pakistan.}

\date{}
\maketitle

\begin{abstract}
This paper develops some new analytical solutions to the
$f(\mathbb{R},\mathbb{T})$ field equations through extended
gravitational decoupling. For this purpose, we take spherical
anisotropic configuration as a seed source and extend it to an
additional source. The modified field equations comprise the impact
of both sources which are then decoupled into two distinct sets by
applying the transformations on $g_{tt}$ and $g_{rr}$ metric
potentials. The original anisotropic source is adorned by the first
sector, and we make it solvable by considering two different
well-behaved solutions. The second sector is in terms of an
additional source and we adopt some constraints to find deformation
functions. The first constraint is the isotropization condition
which transforms the total fluid distribution into an isotropic
system only for a specific value of the decoupling parameter. The
other constraint is taken as the complexity-free fluid distribution.
The unknown constants are calculated at the hypersurface through
matching conditions. The preliminary information (mass and radius)
of a compact star $4U 1820-30$ is employed to analyze physical
attributes of the resulting models. We conclude that certain values
of both the coupling as well as decoupling parameter yield viable
and stable solutions in this theory.
\end{abstract}

\section{Introduction}

The well-structured but perplexing composition of our cosmos
includes massive geometrical bodies such as stars, clusters and
other mysterious constituents. Since the universe has passed through
different epochs, its evolution has become a topic of great interest
for astrophysicists. The cosmos, in which we are living, is going
through a very special era, i.e., it is currently expanding with an
acceleration rate. Cosmologists confirmed such expansion by
performing several attempts on distant galaxies. It is claimed that
an abundance of a mysterious force with large repulsive pressure
(known as the dark energy) causes this rapid expansion. General
relativity ($\mathbb{GR}$) is the first ever theory which helps
researchers to figure out both cosmological and astrophysical
phenomena properly. However, this theory faces two major drawbacks
(cosmic coincidence and fine-tuning), and thus its multiple
extensions have recently been introduced to tackle with these
issues. The $f(\mathbb{R})$ theory is the simplest extension to
$\mathbb{GR}$ which was obtained by putting generic function of the
Ricci scalar in place of $\mathbb{R}$ in an Einstein-Hilbert action.
Various authors \cite{2,2a} discussed this theory to explain
inflationary as well as present eras. The regions in which multiple
$f(\mathbb{R})$ models show stable behavior have been discussed with
the help of different techniques \cite{9,9g}.

The concept of matter-geometry interaction in $f(\mathbb{R})$
framework was introduced by Bertolami et al. \cite{10} for the very
first time. They added the role of geometrical quantity $\mathbb{R}$
in the Lagrangian $\mathbb{L}_{m}$ and analyzed how physical
features of celestial bodies vary. After this, many people put their
attention to generalize such couplings at some action level which
may help in figuring out cosmic accelerated expansion. Harko et al.
\cite{20} developed $f(\mathbb{R},\mathbb{T})$ theory, in which
$\mathbb{T}$ expresses trace of the energy-momentum tensor
($\mathbb{EMT}$) whose inclusion yields new gravitational aspects.
The role of minimal as well as non-minimal interaction on test
particles can be studied through different models of this gravity.
There exists an additional force (depends on density and pressure
\cite{22}) due to the non-conserved nature of the corresponding
$\mathbb{EMT}$ in this theory. A minimal $f(\mathbb{R},\mathbb{T})$
model has been employed to explain switching of the matter-dominated
epoch into late-time acceleration era \cite{22a}. A particular
$\mathbb{R}+2\varpi\mathbb{T}$ model has widely been used by
researchers through which several noteworthy results are obtained.
Das et al. \cite{22b} studied gravastar model with three different
equations of state, representing its inner, intermediate and outer
regions in this theory. Deb et al. \cite{25} explored strange star
with isotropic distribution and obtained acceptable behavior of the
developed solution. Various authors \cite{25a,25b} labeled this
theory as the best approach to discuss stellar interiors.

The non-linear nature of the field equations describing a
self-gravitating object makes it much difficult to solve them
analytically. It has always been interesting to calculate
well-behaved solutions of such complicated equations and study
physical feasibility of realistic systems. In this regard, a large
body of literature is stuffed with solutions which have been
formulated through multiple techniques. The gravitational decoupling
is one of the recently developed techniques to formulate feasible
solutions analogous to the fluid distribution involving multiple
sources (such as anisotropy, dissipation flux and shear). The
minimal geometric deformation (MGD) was firstly developed by Ovalle
\cite{32} in the context of braneworld to find exact solutions
representing compact models. Later on, Ovalle et al. \cite{33}
employed this strategy to calculate extended solutions from the seed
isotropic domain and observed their nature through graphical
analysis. Sharif and Sadiq \cite{34} determined two exact solutions
corresponding to charged sphere in $\mathbb{GR}$ through which the
influence of electromagnetic field was observed on them. An
isotropic solution was extended into multiple anisotropic models in
$f(\mathbb{G})$ and $f(\mathbb{R})$ frameworks \cite{35}.

Estrada and Tello-Ortiz \cite{36a} developed some stable models by
extending the domain of isotropic Heintzmann solution in this
framework. Hensh and Stuchlik \cite{37a} found that modified Tolman
VII solution entails pressure anisotropy on which the influence of
decoupling parameter is also observed. The MGD approach has proved
to be successful in obtaining analytical solutions, however it has a
drawback, i.e., it does not explain a black hole structure with a
well-defined horizon. As a result, Ovalle \cite{26} deformed both
the metric components, and thus named it as the extended geometric
deformation (EGD). Contreras and Bargue\~{n}o \cite{36} assumed
vacuum BTZ ansatz and extended it to exterior charged BTZ solution
by means of EGD. Sharif and his collaborators \cite{37c} constructed
multiple acceptable extensions in $\mathbb{GR}$, $f(\mathbb{R})$ and
Brans-Dicke theories. Zubair et al. \cite{37d} formulated the
extended charged Heintzmann solutions and found them stable for
specific values of the decoupling parameter. We have analyzed the
role of matter-geometry coupling on several anisotropic compact
models \cite{37f}.

The notion of complexity in self-gravitating systems has been
recognized as an interesting topic in the last few years.
Researchers developed multiple definitions in such a way that the
parameters defining complexity of the physical structure can
completely describe that system but found them inadequate. In this
regard, the most appropriate definition was recently established for
the case of static sphere \cite{37g} and then extended to the
dynamical matter source \cite{37h}. It was found that the orthogonal
decomposition of the curvature tensor yields some scalars
incorporating different physical attributes such as the
inhomogeneous energy density and local anisotropy. This idea has
been extended for static/dynamical configurations in modified
framework \cite{37i,37ia}. The complexity-free condition can be used
as a constraint to solve the complicated field equations. This
constraint has been used to formulate acceptable solutions through
decoupling scheme \cite{37j}. Casadio et al. \cite{37k} applied MGD
strategy on the field equations and calculated their solutions
corresponding to the isotropization as well as complexity-free
conditions. Maurya et al. \cite{37l} explored the impact of
decoupling parameters on both these solutions with the help of MGD
and EGD. Sharif and Majid \cite{37m} adopted two different forms of
metric components to extend this work in the context of Brans-Dicke
theory and found physically feasible models.

This paper explores the effect of $f(\mathbb{R},\mathbb{T})$ gravity
on two different decoupled solutions representing static spherically
symmetric interior distribution. We firstly isotropize the system by
considering the vanishing anisotropy for $\beta=1$, and obtain the
corresponding solution. We then formulate the other solution by
using the complexity dependent constraint, i.e., total matter source
with zero complexity. The paper has the following format. In the
next section, we discuss some basic quantities and the modified
field equations influenced from additional source. We then apply EGD
scheme to decouple these equations in section \textbf{3}. We
formulate two solutions corresponding to the above constraints in
sections \textbf{4} and \textbf{5}. The graphical interpretation of
physical features is given in section \textbf{6}. Finally, we
conclude our results in section \textbf{7}.

\section{The $f(\mathbb{R},\mathbb{T})$ Theory}

The inclusion of an additional field along with the modified
functional in the Einstein-Hilbert action (with $\kappa=8\pi$)
yields \cite{20}
\begin{equation}\label{g1}
S_{f(\mathbb{R},\mathbb{T})}=\int\frac{1}{16\pi}\left[f(\mathbb{R},\mathbb{T})
+\mathbb{L}_{m}+\beta\mathbb{L}_{\mathfrak{B}}\right]\sqrt{-g}d^{4}x,
\end{equation}
where the Lagrangian corresponding to the seed and additional
sources are represented by $\mathbb{L}_{m}$ and
$\mathbb{L}_{\mathfrak{B}}$, respectively. Here, the influence of an
extra source (gravitationally coupled to the original source) on
self-gravitating system is controlled by the decoupling parameter
$\beta$. The variation of the action \eqref{g1} with respect to
$g_{\eta\zeta}$ provides the field equations in modified gravity as
\begin{equation}\label{g2}
\mathbb{G}_{\eta\zeta}=8\pi \mathbb{T}_{\eta\zeta}^{(Tot)},
\end{equation}
where the Einstein tensor $\mathbb{G}_{\eta\zeta}$ describes
geometry of the considered structure whereas the matter
configuration is characterized by right hand side which can further
be classified as
\begin{equation}\label{g3}
\mathbb{T}_{\eta\zeta}^{(Tot)}=\mathbb{T}_{\eta\zeta}^{(Eff)}+\beta\mathfrak{B}_{\eta\zeta}=\frac{1}{f_{\mathbb{R}}}
\mathbb{T}_{\eta\zeta}+\mathbb{T}_{\eta\zeta}^{(D)}+\beta\mathfrak{B}_{\eta\zeta}.
\end{equation}
Here, $\mathfrak{B}_{\eta\zeta}$ is a newly added source and
$\mathbb{T}_{\eta\zeta}^{(Eff)}$ is the effective $\mathbb{EMT}$ of
$f(\mathbb{R},\mathbb{T})$ theory. The modified sector
$\mathbb{T}_{\eta\zeta}^{(D)}$ of this gravity has the form
\begin{eqnarray}\nonumber
\mathbb{T}_{\eta\zeta}^{(D)}&=&\frac{1}{8\pi f_{\mathbb{R}}}
\bigg[f_{\mathbb{T}}\mathbb{T}_{\eta\zeta}+\bigg\{\frac{1}{2}\big(f-\mathbb{R}f_{\mathbb{R}}\big)
-\mathbb{L}_{m}f_{\mathbb{T}}\bigg\}g_{\zeta\eta}\\\label{g4}
&+&\big(\nabla_{\eta}\nabla_{\zeta}-g_{\eta\zeta}\Box\big)f_{\mathbb{R}}+2f_{\mathbb{T}}g^{\rho\alpha}\frac{\partial^2
\mathbb{L}_{m}}{\partial g^{\eta\zeta}\partial
g^{\rho\alpha}}\bigg],
\end{eqnarray}
where $f_{\mathbb{T}}=\frac{\partial
f(\mathbb{R},\mathbb{T})}{\partial \mathbb{T}}$ and
$f_{\mathbb{R}}=\frac{\partial f(\mathbb{R},\mathbb{T})}{\partial
\mathbb{R}}$. Furthermore, $\Box\equiv
\frac{1}{\sqrt{-g}}\partial_\eta\big(\sqrt{-g}g^{\eta\zeta}\partial_{\zeta}\big)$
and $\nabla_\eta$ symbolize the D'Alembert operator and covariant
derivative, respectively.

The nature of the original (seed) fluid distribution is assumed to
be anisotropic which can be characterized by the $\mathbb{EMT}$
given as
\begin{equation}\label{g5}
\mathbb{T}_{\eta\zeta}=(\mu+P_\bot)\mathcal{K}_{\eta}\mathcal{K}_{\zeta}+\left(P_r-P_\bot\right)\mathcal{W}_\eta\mathcal{W}_\zeta
+P_\bot g_{\eta\zeta},
\end{equation}
where $\mu$ is the energy density, $P_r$ and $P_\bot$ are the
radial/tangential pressures. Also, $\mathcal{W}_{\eta}$ and
$\mathcal{K}_{\eta}$ are treated as the four-vector and
four-velocity, respectively. One can determine the trace of
$f(\mathbb{R},\mathbb{T})$ field equations as follows
\begin{align}\nonumber
&3\nabla^{\eta}\nabla_{\eta}f_\mathbb{R}+4f_\mathbb{T}\mathbb{L}_m+\mathbb{R}f_\mathbb{R}-2f-\mathbb{T}\big(f_\mathbb{T}+1\big)
-2g^{\eta\zeta}g^{\rho\alpha}f_\mathbb{T}\frac{\partial^2\mathbb{L}_m}{\partial
g^{\eta\zeta}\partial g^{\rho\alpha}}=0.
\end{align}
The results of $f(\mathbb{R})$ gravity can be recovered for the case
when $f_\mathbb{T}=0$. As this theory involves the interaction
between geometry and matter, thus the covariant divergence of the
corresponding $\mathbb{EMT}$ is non-null. Consequently, there
appears an additional force in the gravitational field of massive
structures. Thus we obtain
\begin{equation}\label{g11}
\nabla^\eta\mathbb{T}_{\eta\zeta}=\frac{f_\mathbb{T}}{8\pi-f_\mathbb{T}}\bigg[\nabla^\eta\Theta_{\eta\zeta}
+(\mathbb{T}_{\eta\zeta}+\Theta_{\eta\zeta})\nabla^\eta\ln{f_\mathbb{T}}
-\frac{8\pi\beta}{f_\mathbb{T}}\nabla^\eta\mathfrak{B}_{\eta\zeta}
-\frac{1}{2}g_{\rho\alpha}\nabla_\zeta\mathbb{T}^{\rho\alpha}\bigg],
\end{equation}
where
$\Theta_{\eta\zeta}=-2\mathbb{T}_{\eta\zeta}+g_{\eta\zeta}\mathbb{L}_m-2g^{\rho\alpha}\frac{\partial^2
\mathbb{L}_{m}}{\partial g^{\eta\zeta}\partial g^{\rho\alpha}}$ and
$\mathbb{L}_{m}$ is considered to be $P=\frac{P_r+2P_\bot}{3}$ in
this case leading to $\frac{\partial^2 \mathbb{L}_{m}}{\partial
g^{\eta\zeta}\partial g^{\rho\alpha}}=0$.

The geometry of spacetime structure contains two regions, namely
inner and outer which are distinguished by the hypersurface
$\Sigma$, thus the interior of static spherical system is defined by
the following metric
\begin{equation}\label{g6}
ds^2=-e^{\chi} dt^2+e^{\sigma} dr^2+r^2d\theta^2+r^2\sin^2\theta
d\vartheta^2,
\end{equation}
where $\chi=\chi(r)$ and $\sigma=\sigma(r)$. The terms
$\mathcal{W}^\eta$ and $\mathcal{K}^\eta$ for this line element are
calculated as
\begin{equation}\label{g7}
\mathcal{W}^\eta=(0,e^{\frac{-\sigma}{2}},0,0), \quad
\mathcal{K}^\eta=(e^{\frac{-\chi}{2}},0,0,0),
\end{equation}
satisfying the relations
\begin{equation}\nonumber
\mathcal{W}^\eta \mathcal{W}_{\eta}=1, \quad \mathcal{W}^\eta
\mathcal{K}_{\eta}=0, \quad \mathcal{K}^\eta \mathcal{K}_{\eta}=-1.
\end{equation}
We adopt a particular $f(\mathbb{R},\mathbb{T})$ model to make our
results meaningful. The structural transformation of spherical
objects can entirely be discussed by taking the linear model as
follows
\begin{equation}\label{g61}
f(\mathbb{R},\mathbb{T})=f_1(\mathbb{R})+2f_2(\mathbb{T})=\mathbb{R}+2\varpi\mathbb{T},
\end{equation}
with $f_2(\mathbb{T})=\varpi\mathbb{T},~\varpi$ is a real-valued
coupling parameter and $\mathbb{T}=2P_\bot+P_r-\mu$. Houndjo and
Piattella \cite{38} observed that the holographic dark energy
features may be discussed through model \eqref{g61} while exploring
the pressureless fluid configuration. The standard conservation of
the $\mathbb{EMT}$ is also observed to be compatible with this model
\cite{39}.

The field equations analogous to spherical self-gravitating system
\eqref{g6} and modified model \eqref{g61} are
\begin{align}\label{g8}
&e^{-\sigma}\left(\frac{\sigma'}{r}-\frac{1}{r^2}\right)
+\frac{1}{r^2}=8\pi\left(\mu-\beta\mathfrak{B}_{0}^{0}\right)+\frac{\varpi}{3}\big(9\mu-P_r-2P_\bot\big),\\\label{g9}
&e^{-\sigma}\left(\frac{1}{r^2}+\frac{\chi'}{r}\right)
-\frac{1}{r^2}=8\pi\left(P_r+\beta\mathfrak{B}_{1}^{1}\right)-\frac{\varpi}{3}\big(3\mu-7P_r-2P_\bot\big),\\\label{g10}
&\frac{e^{-\sigma}}{4}\left\{2\chi''+\chi'^2-\sigma'\chi'+\frac{2\big(\chi'-\sigma'\big)}{r}\right\}
=8\pi\left(P_\bot+\beta\mathfrak{B}_{2}^{2}\right)-\frac{\varpi}{3}\big(3\mu-P_r-8P_\bot\big),
\end{align}
where the factors multiplied with $\varpi$ appear due to the
modified theory and prime means $\frac{\partial}{\partial r}$.
Moreover, Eq.\eqref{g11} after some manipulation yields
\begin{align}\nonumber
&\frac{dP_r}{dr}+\frac{\chi'}{2}\left(\mu+P_r\right)+\frac{\beta\chi'}{2}
\left(\mathfrak{B}_{1}^{1}-\mathfrak{B}_{0}^{0}\right)+\frac{2}{r}\left(P_r-P_\bot\right)\\\label{g12}
&+\beta\frac{d\mathfrak{B}_{1}^{1}}{dr}+\frac{2\beta}{r}\left(\mathfrak{B}_{1}^{1}
-\mathfrak{B}_{2}^{2}\right)=-\frac{\varpi}{4\pi-\varpi}\big(\mu'-P'\big),
\end{align}
where the term on the right hand side reveals non-conserved nature
of $f(\mathbb{R},\mathbb{T})$ theory. The variations in the
structure of any self-gravitating body can be studied through
Eq.\eqref{g12} which is also identified as the generalized
Tolman-Opphenheimer-Volkoff equation. The unknowns are now increased
as the additional source is incorporated in the field equations.
These unknowns are
$\chi,~\sigma,~\mu,~P_r,~P_\bot,~\mathfrak{B}_{0}^{0},~\mathfrak{B}_{1}^{1}$
and $\mathfrak{B}_{2}^{2}$, thus the only possibility to obtain
analytic solution is the consideration of some constraints. In this
regard, we use an efficient approach \cite{33} to obtain solutions
for the considered setup.

\section{Gravitational Decoupling}

A systematic approach which allows both the temporal as well as
radial metric potentials to be transformed to get solutions of the
field equations is known as the extended gravitational decoupling.
We employ this scheme to decouple the field equations so that we can
solve each sector separately. We assume the following solution of
Eqs.\eqref{g8}-\eqref{g10} to implement this technique as
\begin{equation}\label{g15}
ds^2=-e^{\xi(r)}dt^2+\frac{1}{\rho(r)}dr^2+r^2d\theta^2+r^2\sin^2\theta
d\vartheta^2.
\end{equation}
The linear transformations on temporal and radial metric
coefficients are of the form
\begin{equation}\label{g16}
\xi\rightarrow\chi=\xi+\beta\mathcal{G}, \quad \rho\rightarrow
e^{-\sigma}=\rho+\beta\mathcal{F},
\end{equation}
where $\mathcal{G}=\mathcal{G}(r)$ and $\mathcal{F}=\mathcal{F}(r)$
are the temporal and radial deformation functions, respectively. It
is worth noting that these linear mappings do not disturb the
spherical symmetry. The set representing seed (anisotropic) source
can be separated from the field equations \eqref{g8}-\eqref{g10} by
implementing the transformations \eqref{g16} (for $\beta=0$) as
\begin{align}\label{g18}
&e^{-\sigma}\left(\frac{\sigma'}{r}-\frac{1}{r^2}\right)
+\frac{1}{r^2}=8\pi\mu+\frac{\varpi}{3}\big(9\mu-P_r-2P_\bot\big),\\\label{g19}
&e^{-\sigma}\left(\frac{1}{r^2}+\frac{\chi'}{r}\right)-\frac{1}{r^2}=8\pi
P_r-\frac{\varpi}{3}\big(3\mu-7P_r-2P_\bot\big),\\\label{g20}
&\frac{e^{-\sigma}}{4}\left(\chi'^2-\sigma'\chi'+2\chi''-\frac{2\sigma'}{r}+\frac{2\chi'}{r}\right)
=8\pi P_\bot-\frac{\varpi}{3}\big(3\mu-P_r-8P_\bot\big),
\end{align}
which yield the explicit expressions of the state variables as
\begin{align}\nonumber
\mu&=\frac{e^{-\sigma}}{48r^2(\varpi+2\pi)(\varpi+4\pi)}\big[2\big\{r^2\varpi\chi''
+8r\sigma'(\varpi+3\pi)+8(\varpi+3\pi)\\\label{g18a}
&\times\big(e^{\sigma}-1\big)\big\}+r^2\varpi\chi'^2+r\varpi\chi'\big(4-r\sigma'\big)\big],\\\nonumber
P_r&=\frac{e^{-\sigma}}{48r^2(\varpi+2\pi)(\varpi+4\pi)}\big[2\big\{4r\varpi\sigma'-r^2\varpi\chi''
-8(\varpi+3\pi)\big(e^{\sigma}-1\big)\big\}\\\label{g19a}
&-r^2\varpi\chi'^2+r\sigma'\big(20\varpi+r\varpi\sigma'+48\pi\big)\big],\\\nonumber
P_\bot&=\frac{e^{-\sigma}}{48r^2(\varpi+2\pi)(\varpi+4\pi)}\big[10r^2\varpi\chi''
+(5\varpi+12\pi)r^2\chi'^2+24\pi{r^2}\chi''\\\label{g20a}
&-8\varpi+r\chi'\big\{8(\varpi+3\pi)-(5\varpi+12\pi)r\sigma'\big\}-4r\sigma'(\varpi+6\pi)+8{\varpi}e^{\sigma}\big].
\end{align}

Another set characterizing the role of additional source
($\mathfrak{B}_{\eta\zeta}$) is obtained for $\beta=1$ as
\begin{align}\label{g21}
8\pi\mathfrak{B}_{0}^{0}&=\frac{1}{r}\left(\mathcal{F}'+\frac{\mathcal{F}}{r}\right),\\\label{g22}
8\pi\mathfrak{B}_{1}^{1}&=\frac{\mathcal{F}}{r}\left(\chi'+\frac{1}{r}\right)+\frac{e^{-\sigma}\mathcal{G}'}{r},\\\nonumber
8\pi\mathfrak{B}_{2}^{2}&=\frac{\mathcal{F}}{4}\left(2\chi''+\chi'^2+\frac{2\chi'}{r}\right)
+\frac{\mathcal{F}'}{4}\left(\chi'+\frac{2}{r}\right)\\\label{g23}
&+\frac{e^{-\sigma}}{2}\left(\mathcal{G}''+\chi'\mathcal{G}'+\frac{\beta\mathcal{G}'^2}{2}
+\frac{\mathcal{G}'}{r}-\frac{\sigma'\mathcal{G}'}{2}\right).
\end{align}
The EGD scheme allows the two (seed and newly added) matter
distributions to exchange energy between them. Both of these sources
are not conserved individually but the system becomes conserved by
coupling them into a single framework. A successful partition of the
system \eqref{g8}-\eqref{g10} has now been done into two sets. For
the case of first set, we have three equations
\eqref{g18a}-\eqref{g20a} along with five unknown quantities
($\mu,P_r,P_\bot,\chi,\sigma$), so we will assume a well-behaved
solution to make this system solvable. The other set
\eqref{g21}-\eqref{g23} comprises of five unknowns
($\mathcal{G},\mathcal{F},\mathfrak{B}_{0}^{0},\mathfrak{B}_{1}^{1},\mathfrak{B}_{2}^{2}$),
thus two constraints on $\mathfrak{B}$-sector will be required at
the same time to close the system. We define the effective state
variables as follows
\begin{equation}\label{g13}
\breve{\mu}=\mu-\beta\mathfrak{B}_{0}^{0},\quad
\breve{P}_{r}=P_r+\beta\mathfrak{B}_{1}^{1}, \quad
\breve{P}_{\bot}=P_\bot+\beta\mathfrak{B}_{2}^{2},
\end{equation}
and the total anisotropy is
\begin{equation}\label{g14}
\breve{\Pi}=\breve{P}_{\bot}-\breve{P}_{r}=(P_\bot-P_r)+\beta(\mathfrak{B}_{2}^{2}-\mathfrak{B}_{1}^{1})=\Pi+\Pi_{\mathfrak{B}},
\end{equation}
where $\Pi$ and $\Pi_{\mathfrak{B}}$ are anisotropic factors
corresponding to the seed and new source, respectively.

\section{Isotropization of Compact Sources}

We can observe from Eq.\eqref{g14} that there may be a difference
between anisotropic factors generated by the original source
$\mathbb{T}_{\zeta\eta}$ and the total matter distribution, i.e.,
$\Pi$ and $\breve{\Pi}$, respectively. In this sector, we assume
that the incorporation of new source in the original anisotropic
source makes it an isotropic which means $\breve{\Pi}=0$. This
variation in spherical system is controlled by parameter $\beta$,
since $\beta=0$ specifies anisotropic distribution and $1$
represents an isotropic one. We now discuss the later case in order
to isotropize the system which gives
\begin{equation}\label{g14a}
\Pi_{\mathfrak{B}}=-\Pi \quad \Rightarrow \quad
\mathfrak{B}_{2}^{2}-\mathfrak{B}_{1}^{1}=P_r-P_\bot.
\end{equation}
Casadio et al. \cite{37k} considered a self-gravitating system
comprising anisotropic distribution at some initial time and then
used the constraint \eqref{g14a} to convert it into an isotropic
configuration through gravitational decoupling. As we have mentioned
earlier, we are required to take a solution associated with the
original source to construct our solution, thus we have
\begin{eqnarray}\label{g33}
\chi(r)&=&\ln\bigg\{\mathcal{A}^2\bigg(1+\frac{r^2}{\mathcal{B}^2}\bigg)\bigg\},\\\label{g34}
\rho(r)&=&e^{-\sigma}=\frac{\mathcal{B}^2+r^2}{\mathcal{B}^2+3r^2},\\\label{g35}
\mu&=&\frac{3\mathcal{B}^2\big(3\varpi+8\pi\big)+2r^2\big(5\varpi+12\pi\big)}{4\big(\mathcal{B}^2+3r^2\big)^2
\big(\varpi+4\pi\big)\big(\varpi+2\pi\big)},\\\label{g36}
P_r&=&\frac{\varpi\big(3\mathcal{B}^2+2r^2\big)}{4\big(\mathcal{B}^2+3r^2\big)^2\big(\varpi+4\pi\big)
\big(\varpi+2\pi\big)},\\\label{g37}
P_\bot&=&\frac{3\varpi\mathcal{B}^2+4r^2\big(2{\varpi}+3{\pi}\big)}
{4\big(\mathcal{B}^2+3r^2\big)^2\big(\varpi+4\pi\big)\big(\varpi+2\pi\big)},
\end{eqnarray}
where we employe matching criteria to calculate the unknowns, i.e.,
$\mathcal{A}^2$ and $\mathcal{B}^2$. When a cluster of test
particles moves in circular orbit in any gravitational field, the
above solution may help to determine that field \cite{42a1}. This
form of metric potentials has also been employed in $\mathbb{GR}$ to
develop decoupled solutions \cite{37k}.

Some constraints play a significant role in the study of various
characteristics of self-gravitating bodies, known as junction
conditions that provide solutions at the hypersurface
($\Sigma:r=\mathcal{R}$). To discuss the matching conditions, we
take the most appropriate choice of exterior geometry, i.e., the
Schwarzschild metric. The inclusion of higher-order curvature terms
in $f(\mathbb{R})$ gravity (like the Starobinsky model given by
$\mathbb{R}+\alpha\mathbb{R}^2$, where $\alpha$ is a restricted
constant) make the junction conditions different from that in
$\mathbb{GR}$ \cite{41jaa,41jaaa}. However, for the case of model
\eqref{g61}, the first term represents $\mathbb{GR}$ and the second
entity does not contribute to the current setup. Therefore, the
exterior geometry is taken as that of $\mathbb{GR}$ by
\begin{equation}\label{g25}
ds^2=-\frac{r-2\breve{\mathcal{M}}}{r}dt^2+\frac{r}{r-2\breve{\mathcal{M}}}dr^2+
r^2d\theta^2+r^2\sin^2\theta d\vartheta^2,
\end{equation}
where $\breve{\mathcal{M}}$ is the total mass. This provides after
matching with interior spherical geometry \eqref{g6} as
\begin{eqnarray}\label{g37}
\mathcal{A}^2&=&\frac{\mathcal{R}-2\breve{\mathcal{M}}}{\frac{\breve{\mathcal{M}}\mathcal{R}}{\mathcal{R}
-3\breve{\mathcal{M}}}+\mathcal{R}},\\\label{g38}
\mathcal{B}^2&=&\frac{\mathcal{R}^2\big(\mathcal{R}-3\breve{\mathcal{M}}\big)}{\breve{\mathcal{M}}}.
\end{eqnarray}
Further, the graphical interpretation of different physical features
will be studied by using mass ($\breve{\mathcal{M}}=1.58 \pm 0.06
M_{\bigodot}$) and radius ($\mathcal{R}=9.1 \pm 0.4 km$) of $4U
1820-30$ star model \cite{42aa}. Using the constraint \eqref{g14a}
and modified field equations, we obtain a non-linear differential
equation in terms of metric functions \eqref{g33}-\eqref{g34} as
\begin{align}\nonumber
&r\big(\mathcal{B}^2+r^2\big)\big[\big(\mathcal{B}^2+r^2\big)\big\{2r\big((\varpi+4\pi)
\big(\mathcal{B}^4+4\mathcal{B}^2r^2+3r^4\big)\mathcal{G}''(r)+24\pi
r^2\big)\\\nonumber
&+(\varpi+4\pi)r\big(\mathcal{B}^4+4\mathcal{B}^2r^2+3r^4\big)\mathcal{G}'(r)^2-2
(\varpi+4\pi)\big(\mathcal{B}^4+4\mathcal{B}^2r^2-3r^4\big)\\\nonumber
&\times\mathcal{G}'(r)\big\}+2
(\varpi+4\pi)\big(\mathcal{B}^2+2r^2\big)\big(\mathcal{B}^2+3r^2\big)^2
\mathcal{F}'(r)\big]-4\big(\mathcal{B}^2+3r^2\big)^2\\\label{g39}
&\times(\varpi+4\pi)\big(\mathcal{B}^4+2\mathcal{B}^2r^2+2r^4\big)\mathcal{F}(r)=0.
\end{align}
This is the first and second order in $\mathcal{F}(r)$ and
$\mathcal{G}(r)$, respectively. To solve this equation for
$\mathcal{F}(r)$ is an easy task, thus we take $\mathcal{G}(r)=gr^2$
with $g$ as a constant. After substituting this in Eq.\eqref{g39},
the analytic solution takes the form
\begin{align}\nonumber
\mathcal{F}(r)&=\frac{r^2\big(\mathcal{B}^2+r^2\big)}{\mathcal{B}^2+2r^2}\bigg[\mathbb{C}_1
-\frac{1}{9(\varpi+4\pi)}\bigg\{(\varpi+4\pi)g^2\big(\mathcal{B}^2+3r^2\big)+2g\\\label{g40}
&\times(\varpi+4\pi)(\mathcal{B}^2g+3)\ln\big(\mathcal{B}^2+3r^2\big)
+\frac{6(\varpi\mathcal{B}^2g+\pi(4\mathcal{B}^2g-6))}{\mathcal{B}^2+3r^2}\bigg\}\bigg],
\end{align}
where $\mathbb{C}_1$ is an integration constant.

The deformed temporal and radial metric components \eqref{g16} now
become
\begin{align}\label{g40a}
e^{\chi}&=\mathcal{A}^2e^{\beta{gr^2}}\bigg(1+\frac{r^2}{\mathcal{B}^2}\bigg),\\\nonumber
e^{\sigma}&=\rho^{-1}=9\big(\mathcal{B}^2+2r^2\big)\big(\mathcal{B}^2+3r^2\big)\bigg[\beta{r^2}\big(\mathcal{B}^2+r^2\big)
\bigg\{9\mathbb{C}_1\big(\mathcal{B}^2+3r^2\big)\\\nonumber
&-g^2\big(\mathcal{B}^2+3r^2\big)^2-2g\big(\mathcal{B}^2+3r^2\big)(\mathcal{B}^2g+3)
\ln\big(\mathcal{B}^2+3r^2\big)-6\mathcal{B}^2g\\\label{g40aa}
&+\frac{36\pi}{\varpi+4\pi}\bigg\}+9\big(\mathcal{B}^2+r^2\big)\big(\mathcal{B}^2+2r^2\big)\bigg]^{-1}.
\end{align}
Hence, the extended gravitationally decoupled solution representing
the sphere \eqref{g8}-\eqref{g10} is given by the following metric
as
\begin{align}\nonumber
ds^2&=-\mathcal{A}^2e^{\beta{gr^2}}\bigg(1+\frac{r^2}{\mathcal{B}^2}\bigg)dt^2+\frac{\mathcal{B}^2+3r^2}
{\mathcal{B}^2+r^2+\beta\mathcal{F}\big(\mathcal{B}^2+3r^2\big)}dr^2\\\label{g41}
&+r^2d\theta^2+r^2\sin^2\theta d\vartheta^2.
\end{align}
The resulting state variables (the energy density and
radial/tangential pressures) are
\begin{align}\nonumber
\breve{\mu}&=\frac{1}{72\big(\varpi+4\pi\big)
\big(\mathcal{B}^2+3r^2\big)^2}\bigg[\frac{18\varpi\big(9\mathcal{B}^2+10r^2\big)
+432\pi\big(\mathcal{B}^2+r^2\big)}{\varpi+2\pi}\\\nonumber
&+\frac{\beta}{\pi\big(\mathcal{B}^2+2r^2\big)^2}\bigg\{3 (\varpi +4
\pi ) \mathcal{B}^{10} g^2+(\varpi +4 \pi ) \mathcal{B}^8 \big(2 g
\big(26 r^2 g+9\big)\\\nonumber &-27 \mathbb{C}_1\big)+3
\mathcal{B}^6 \big(\varpi r^2 \big(4 g \big(23 r^2 g+8\big)-75
\mathbb{C}_1\big)-4 \pi \big(r^2 \big(75 \mathbb{C}_1-4 g
\\\nonumber &\times\big(23
r^2 g+8\big)\big)+9\big)\big)+45 \mathcal{B}^4 r^2 \big(\varpi  r^2
\big(2 g \big(7 r^2 g+3\big)-15 \mathbb{C}_1\big)+4 \pi\\\nonumber
&\times\big(r^2 \big(2 g \big(7 r^2 g+3\big)-15
\mathbb{C}_1\big)-2\big)\big)+2 (\varpi +4 \pi ) g
\big(\mathcal{B}^2+3 r^2\big)^2 \\\nonumber &\times\big(3
\mathcal{B}^4+7 \mathcal{B}^2 r^2+6 r^4\big) (\mathcal{B}^2g+3) \ln
\big(\mathcal{B}^2+3 r^2\big)+9 \mathcal{B}^2 r^4 \big(\varpi r^2
\big(g\\\nonumber &\times \big(73 r^2 g+48\big)-99
\mathbb{C}_1\big)+4 \pi \big(r^2 \big(g \big(73 r^2 g+48\big)-99
\mathbb{C}_1\big)-9\big)\big)+54 r^6\\\label{g46} &\times
\big(\varpi r^2 \big(g \big(5 r^2 g+4\big)-9 \mathbb{C}_1\big)+4 \pi
\big(r^2 \big(g \big(5 r^2 g+4\big)-9
\mathbb{C}_1\big)-1\big)\big)\bigg\}\bigg],\\\nonumber
\breve{P}_{r}&=\frac{1}{72\pi}\bigg[\frac{18\pi\varpi(3\mathcal{B}^2+2r^2)}{\big(\varpi+2\pi\big)\big(\varpi+4\pi\big)
\big(\mathcal{B}^2+3r^2\big)^2}+\frac{\beta}{\big(\mathcal{B}^2+2r^2\big)\big(\mathcal{B}^2+3r^2\big)}\\\nonumber
&\times\bigg\{\big(\mathcal{B}^2+3 r^2\big) \bigg(\frac{36 \pi
}{\varpi +4 \pi }+9 \mathbb{C}_1 \big(\mathcal{B}^2+3
r^2\big)-g^2\big(\mathcal{B}^2+3 r^2\big)^2-2g\\\nonumber &\times
\big(\mathcal{B}^2+3 r^2\big) (\mathcal{B}^2 g+3) \ln
\big(\mathcal{B}^2+3 r^2\big)-6 \mathcal{B}^2g\bigg)+18 g
\big(\mathcal{B}^2+r^2\big)\\\label{g47} & \big(\mathcal{B}^2+2
r^2\big)\bigg\}\bigg],
\\\nonumber \breve{P}_{\bot}&=-\frac{1}{72\pi\big(\mathcal{B}^2+2r^2\big)\big(\mathcal{B}^2+3r^2\big)^2
\big(\varpi+2\pi\big)\big(\varpi+4\pi\big)}\bigg[\beta(\varpi +2 \pi
) \\\nonumber &\times\big(\varpi  \big(\mathcal{B}^2+3 r^2\big)
\big(g^2 \big(\mathcal{B}^2+3 r^2\big)^3-12 g \big(\mathcal{B}^4+3
\mathcal{B}^2 r^2+3 r^4\big)-9 \mathbb{C}_1\\\nonumber &\times
\big(\mathcal{B}^2+3 r^2\big)^2\big)+4 \pi \big(-9
\big(\mathcal{B}^4 (\mathcal{B}^2\mathbb{C}_1+1)+3 r^4 (9
\mathcal{B}^2 \mathbb{C}_1+1)\\\nonumber &+3 \mathcal{B}^2 r^2 (3
\mathcal{B}^2 \mathbb{C}_1+1)+27 \mathbb{C}_1 r^6\big)-12 g
\big(\mathcal{B}^4+3 \mathcal{B}^2 r^2+3 r^4\big)
\big(\mathcal{B}^2+3 r^2\big)\\\nonumber &+g^2 \big(\mathcal{B}^2+3
r^2\big)^4\big)\big)-18 \pi \varpi \big(\mathcal{B}^2+2 r^2\big)
\big(3 \mathcal{B}^2+2 r^2\big)+2\beta  g (\varpi +2 \pi )
\\\label{g48} &\times (\varpi
+4 \pi )\big(\mathcal{B}^2+3 r^2\big)^3 (\mathcal{B}^2g+3) \ln
\big(\mathcal{B}^2+3 r^2\big)\bigg].
\end{align}
The pressure anisotropy in this case becomes
\begin{eqnarray}\label{g49}
\breve{\Pi}&=&\frac{3r^2\big(1-\beta\big)}{2\big(\varpi+4\pi\big)
\big(\mathcal{B}^2+3r^2\big)^2},
\end{eqnarray}
which vanishes on account of $\beta=1$. These equations represent
our first analytic solution to the modified field equations for
$\beta\in[0,1]$. It is observed that the original anisotropic system
is now transformed into an isotropic system (for $\beta=1$).
Consequently, the change in $\beta$ from $0$ to $1$ follows the
process of isotropization of the considered setup.

\section{Complexity of Compact Sources}

In this section, we are dealing with the complexity of static
spherical source which was recently presented by Herrera \cite{37g}.
Later, this definition was also extended to the case of non-static
dissipative scenario \cite{37h}. The fundamental notion of this
definition is that the structure coupled with uniform/isoropic fluid
distribution considered to be the complexity-free. On the other
hand, some scalars can be calculated from orthogonal splitting of
the Riemann tensor, one of them is $\mathbb{Y}_{TF}$ that involves
inhomogenious/anisotropic factors, and thus adopted as the
complexity factor. We extend Herrera's definition \cite{37g} to
$f(\mathbb{R},\mathbb{T})$ theory to explore the effects of modified
corrections on the complexity. Consequently, $\mathbb{Y}_{TF}$
becomes
\begin{equation}\label{g51}
\mathbb{Y}_{TF}(r)=-\frac{4\pi}{r^3}\int_0^ry^3\mu'(y)dy+8\pi\Pi\big(\varpi+1\big),
\end{equation}
and the Tolman mass for this scenario can be specified as
\begin{equation}\label{g52}
m_T=4\pi\int_0^ry^2e^{\frac{(\chi+\sigma)}{2}}\big(\mu+3P\big)dy,
\end{equation}
which helps in determining the sphere's total energy. Combining the
complexity factor \eqref{g51} with Tolman mass, we obtain
\begin{equation}\label{g53}
m_T=\breve{\mathcal{M}}_T\left(\frac{r}{\mathcal{R}}\right)^3+r^3\int_r^{\mathcal{R}}\frac{e^{\frac{(\chi+\sigma)}{2}}}{y}
\big(\mathbb{Y}_{TF}-8\pi\Pi\varpi\big)dy,
\end{equation}
where $\breve{\mathcal{M}}_T$ indicates the Tolman mass at the
hypersurface.

As we have coupled two different sources, thus the factor
$\breve{\mathbb{Y}}_{TF}$ corresponding to the total distribution
\eqref{g8}-\eqref{g10} comes out to be
\begin{eqnarray}\nonumber
\breve{\mathbb{Y}}_{TF}(r)&=&8\pi\tilde{\Pi}\big(\varpi+1\big)-\frac{4\pi}{r^3}\int_0^ry^3\breve{\mu}'(y)dy\\\nonumber
&=&8\pi\Pi\big(\varpi+1\big)-\frac{4\pi}{r^3}\int_0^ry^3\mu'(y)dy\\\label{g54}
&+&8\pi\Pi_{\mathfrak{B}}\big(\varpi+1\big)+\frac{4\pi}{r^3}\int_0^ry^3\mathfrak{B}{_0^0}'(y)dy,
\end{eqnarray}
whose compact form is
\begin{eqnarray}\label{g55}
\breve{\mathbb{Y}}_{TF}=\mathbb{Y}_{TF}+\mathbb{Y}_{TF}^{\mathfrak{B}},
\end{eqnarray}
where $\mathbb{Y}_{TF}$ and $\mathbb{Y}_{TF}^{\mathfrak{B}}$
associate with the seed \eqref{g18a}-\eqref{g20a} and additional
source \eqref{g21}-\eqref{g23}, respectively. The solution
\eqref{g46}-\eqref{g49} is established for the constraint
$\breve{\Pi}=0$ that assists Eq.\eqref{g54} to yield
\begin{eqnarray}\label{g56}
\breve{\mathbb{Y}}_{TF}&=&-\frac{4\pi}{r^3}\int_0^ry^3\breve{\mu}'(y)dy,
\end{eqnarray}
representing the complexity factor of geometry \eqref{g41}. Its
value in terms of the field equations is casted as
\begin{align}\nonumber
\breve{\mathbb{Y}}_{TF}&=-\frac{1}{216r^3\big(\varpi+2\pi\big)\big(\varpi+4\pi\big)}
\bigg[\frac{1}{\big(\mathcal{B}^4+5\mathcal{B}^2r^2
+6r^4\big)^2}\bigg\{48(\varpi +2 \pi ) \beta  r^5 \\\nonumber
&\times\big(4 \pi \big(\mathcal{B}^4 (\mathcal{B}^2 g (4
\mathcal{B}^2 g-3)+18)+9 g r^6 (7 \mathcal{B}^2 g+6)+9 r^4
\big(\mathcal{B}^2 g (7 \mathcal{B}^2 g+9)\\\nonumber &+3\big)+9
\mathcal{B}^2 r^2 (\mathcal{B}^2 g (3 \mathcal{B}^2 g+2)+6)+27 g^2
r^8\big)+\varpi g \big(\mathcal{B}^6 (4 \mathcal{B}^2 g-3)+9
\mathcal{B}^4 \\\nonumber &\times r^2 (3 \mathcal{B}^2 g+2)+9 r^6 (7
\mathcal{B}^2 g+6)+9 \mathcal{B}^2 r^4 (7 \mathcal{B}^2 g+9)+27 g
r^8\big)-\mathcal{B}^2 g \\\nonumber &\times (\varpi +4 \pi )
(\mathcal{B}^2 g+3) \big(\mathcal{B}^2+3 r^2\big)^2\ln
\big(\mathcal{B}^2+3 r^2\big)\big)-36 \pi \big(\mathcal{B}^2+2
r^2\big)^2 \\\nonumber &\times\bigg(3 \varpi r \big(\mathcal{B}^4+5
\mathcal{B}^2 r^2+40 r^4\big)-\sqrt{3} \varpi \sqrt{\mathcal{B}^2}
\big(\mathcal{B}^2+3 r^2\big)^2 \tan ^{-1}\bigg(\frac{\sqrt{3}
r}{\sqrt{\mathcal{B}^2}}\bigg)\\\label{g56a} &+288 \pi
r^5\bigg)\bigg\}+\frac{216 (\varpi+2 \pi ) (\varpi+4 \pi )
\mathcal{B}^2 \beta \mathbb{C}_1 r^5}{\big(\mathcal{B}^2+2
r^2\big)^2}\bigg].
\end{align}

\subsection{Two Systems possessing the Same Complexity Factor}

Here, the newly added source $\mathfrak{B}_{\eta\zeta}$ is assumed
to be complexity-free, i.e., $\mathbb{Y}_{TF}^{\mathfrak{B}}=0$,
thus its insertion in the seed source does not affect the
corresponding $\mathbb{Y}_{TF}$ and hence leading to
$\breve{\mathbb{Y}}_{TF}=\mathbb{Y}_{TF}$ or
\begin{eqnarray}\label{g56b}
8\pi\Pi_{\mathfrak{B}}\big(\varpi+1\big)=-\frac{4\pi}{r^3}\int_0^ry^3\mathfrak{B}{_0^0}'(y)dy.
\end{eqnarray}
The right side together with Eq.\eqref{g21} provides
\begin{eqnarray}\label{g56c}
-\frac{4\pi}{r^3}\int_0^ry^3\mathfrak{B}{_0^0}'(y)dy=\frac{1}{r^2}\bigg(\mathcal{F}-\frac{r\mathcal{F}'}{2}\bigg),
\end{eqnarray}
which after substitution in Eq.\eqref{g56b} yields the differential
equation as
\begin{eqnarray}\nonumber
&&(\varpi+1)\bigg\{\mathcal{F}'(r)\bigg(\frac{\chi'}{4}+\frac{1}{2r}\bigg)+\mathcal{F}(r)\bigg(\frac{\chi''}{2}-\frac{1}{r^2}
+\frac{\chi'^2}{4}-\frac{\chi'}{2r}\bigg)\\\nonumber
&&+e^{-\sigma}\bigg(\frac{\mathcal{G}''(r)}{2}+\frac{\mathcal{G}'(r)\chi'}{2}-\frac{\mathcal{G}'(r)
\sigma'}{4}+\frac{\beta\mathcal{G}'(r)^2}{4}-\frac{\mathcal{G}'(r)}{2r}\bigg)\bigg\}\\\label{g56d}
&&-\frac{1}{2}\bigg(\frac{2\mathcal{F}(r)}{r^2}-\frac{\mathcal{F}'(r)}{r}\bigg)=0.
\end{eqnarray}
This is in terms of both the deformation functions, thus we again
take $\mathcal{G}(r)=gr^2$ as used earlier to solve this for
$\mathcal{F}(r)$. Equation \eqref{g56d} depends on unknown metric
components describing the original source $\mathbb{T}_{\eta\zeta}$,
thus we need to employ the Tolman IV ansatz to determine the
solution. These components are
\begin{align}\label{g57}
\chi(r)&=\ln\bigg\{\mathcal{A}^2\bigg(1+\frac{r^2}{\mathcal{B}^2}\bigg)\bigg\},\\\label{g58}
\rho(r)&=e^{-\sigma}=\frac{\big(\mathcal{B}^2+r^2\big)\big(\mathcal{C}^2-r^2\big)}{\mathcal{C}^2\big(\mathcal{B}^2+2r^2\big)},
\end{align}
which produce the energy density \eqref{g18a} and isotropic pressure
($P_r$ and $P_{\bot}$ are equal in this case) as
\begin{align}\nonumber
\mu&=\frac{1}{4\mathcal{C}^2\big(\mathcal{B}^2+2r^2\big)^2\big(\varpi+2\pi\big)\big(\varpi+4\pi\big)}
\big[4\big(\varpi+3\pi\big)\mathcal{B}^4+\mathcal{B}^2\\\nonumber
&\times\big(8\varpi{r^2}+5\varpi\mathcal{C}^2+28\pi{r^2}+12\pi\mathcal{C}^2\big)
+2r^2\big\{4\pi\big(3r^2+\mathcal{C}^2\big)\\\label{g59}
&+\varpi\big(3r^2+2\mathcal{C}^2\big)\big\}\big],\\\nonumber
P&=\frac{1}{4\mathcal{C}^2\big(\mathcal{B}^2+2r^2\big)^2\big(\varpi+2\pi\big)\big(\varpi+4\pi\big)}
\big[\mathcal{B}^2\big\{4\pi\big(\mathcal{C}^2-5r^2\big)-4\varpi{r^2}\\\label{g60}
&+3\varpi\mathcal{C}^2\big\}-4\pi\mathcal{B}^4+2r^2\big\{2\varpi\mathcal{C}^2
-3\varpi{r^2}-4\pi\big(3r^2-\mathcal{C}^2\big)\big\}\big].
\end{align}

Equations \eqref{g37} and \eqref{g38} provide two constants
$\mathcal{A}^2$ and $\mathcal{B}^2$, while we have $\mathcal{C}^2$
as
\begin{align}\label{g60a}
\mathcal{C}^2&=\frac{\mathcal{R}^3}{\breve{\mathcal{M}}}.
\end{align}
Substituting the metric potentials \eqref{g57} and \eqref{g58} in
\eqref{g56d}, we have
\begin{align}\nonumber
\mathcal{F}(r)&=\frac{r^2\big(\mathcal{B}^2+r^2\big)}{\mathcal{B}^2\big(2+\varpi\big)+r^2\big(2\varpi+3\big)}\bigg[\mathbb{C}_2
-\frac{g\big(1+\varpi\big)}{8\mathcal{C}^2}\bigg\{\big(\mathcal{B}^2+2\mathcal{C}^2\big)\\\nonumber
&\times(\mathcal{B}^2\beta{g}+4)\ln\big(\mathcal{B}^2+2r^2\big)
-2r^2\big(\mathcal{B}^2\beta{g}-2\beta\mathcal{C}^2g+6\big)\\\label{g60b}
&+\frac{2\mathcal{B}^2(\mathcal{B}^2+2\mathcal{C}^2)}{\mathcal{B}^2+2r^2}-2\beta{g}r^4\bigg\}\bigg],
\end{align}
where $\mathbb{C}_2$ identifies as an integration constant. Hence,
the temporal metric function is the same as given in Eq.\eqref{g40a}
whereas the radial component has the form
\begin{align}\nonumber
e^{\sigma}&=\rho^{-1}=8\mathcal{C}^2\big\{\big(\varpi+2\big)\mathcal{B}^2+\big(2\varpi+3\big)r^2\big\}
\bigg[8\big(\mathcal{C}^2-r^2\big)\big\{\big(\varpi+2\big)\mathcal{B}^2\\\nonumber
&+\big(2\varpi+3\big)r^2\big\}+\beta{r^2}\big(\mathcal{B}^2+r^2\big)\bigg\{8\mathcal{C}^2\mathbb{C}_2+g\big(1+\varpi)
\bigg(2r^2\big(\beta{g}\\\nonumber
&\times\big(\mathcal{B}^2-2\mathcal{C}^2\big)+6\big)-\big(\mathcal{B}^2+2\mathcal{C}^2\big)
\big(\beta\mathcal{B}^2g+4\big)\ln\big(\mathcal{B}^2+2r^2\big)\\\label{g60c}
&+2\beta{g}r^4-\frac{2\mathcal{B}^2\big(\mathcal{B}^2+2\mathcal{C}^2\big)}{\mathcal{B}^2+2r^2}\bigg)\bigg\}\bigg]^{-1}.
\end{align}
Equation \eqref{g54} offers the definition of
$\breve{\mathbb{Y}}_{TF}$ which in the current setup leads to
\begin{align}\nonumber
\breve{\mathbb{Y}}_{TF}&=\mathbb{Y}_{TF}=\frac{\pi\big(\mathcal{B}^2+2\mathcal{C}^2\big)}{16r^3\mathcal{C}^2\big(\mathcal{B}^2
+2r^2\big)^2\big(\varpi+2\pi\big)\big(\varpi+4\pi\big)}\bigg[6\varpi\mathcal{B}^4r+128\pi{r^5}\\\label{g60d}
&-3\varpi\sqrt{2\mathcal{B}^2}\big(\mathcal{B}^2+2r^2\big)^2\tan^{-1}\bigg(r\sqrt{\frac{2}{\mathcal{B}^2}}\bigg)
+64\varpi{r^5}+20\varpi\mathcal{B}^2r^3\bigg].
\end{align}

\subsection{Generating Solutions with Zero Complexity}

We now find the other solution by discussing the case in which the
original source with complexity $\mathbb{Y}_{TF}\neq0$ becomes free
from complexity after the inclusion of an additional source.
Therefore, the total fluid configuration has no complexity, i.e.,
$\breve{\mathbb{Y}}_{TF}=0$, which makes Eq.\eqref{g55} in terms of
Tolman IV components and $\mathcal{G}(r)=gr^2$ as
\begin{align}\nonumber
&8r\big\{r\mathcal{F}'(r)-2\mathcal{F}(r)\big\}+\frac{1}{\mathcal{C}^2\big(\mathcal{B}^2+r^2\big)^2\big(\mathcal{B}^2+2r^2\big)^2
\big(\varpi+2\pi\big)\big(\varpi+4\pi\big)}\\\nonumber &\bigg[\pi
\big(\mathcal{B}^2+r^2\big)^2 (\mathcal{B}^2+2 \mathcal{C}^2)
\bigg\{\varpi  \bigg(6 \mathcal{B}^4 r+20 \mathcal{B}^2 r^3-3
\sqrt{2\mathcal{B}^2} \big(\mathcal{B}^2+2 r^2\big)^2\\\nonumber
&\times \tan ^{-1}\bigg(\frac{\sqrt{2}
r}{\sqrt{\mathcal{B}^2}}\bigg)+64 r^5\bigg)+128 \pi r^5\bigg\}-8
(\varpi+4\pi )(\varpi +2\pi )(\varpi +1)\\\nonumber &\times  r
\big\{\mathcal{C}^2 \big(\mathcal{B}^2+2 r^2\big)^2 \big(2
\big(\mathcal{B}^4+2 \mathcal{B}^2 r^2+2 r^4\big)
\mathcal{F}(r)-r\mathcal{F}'(r)\big(\mathcal{B}^2+r^2\big)\\\nonumber
&\times \big(\mathcal{B}^2+2 r^2\big) \big)+2 r^4 g
\big(\mathcal{B}^2+r^2\big)^2 \big(\beta
g\big(\mathcal{B}^2+r^2\big) \big(\mathcal{B}^2+2 r^2\big)
\big(r^2-\mathcal{C}^2\big)\\\label{g60e} &+\mathcal{B}^4+4
\mathcal{B}^2 r^2-\mathcal{B}^2 \mathcal{C}^2+6 r^4-4 r^2
\mathcal{C}^2\big)\big\}\bigg]=0,
\end{align}
which is the first order differential equation whose solution is
\begin{align}\nonumber
\mathcal{F}(r)&=\frac{r^2\big(\mathcal{B}^2+r^2\big)}{8\mathcal{C}^2\big(\varpi^2+6\pi\varpi+8\pi^2\big)
\big(\varpi\mathcal{B}^2+2\varpi{r^2}+2\mathcal{B}^2+3r^2\big)}\bigg[\big(2\pi+\varpi\big)\\\nonumber
&\times\big(4\pi+\varpi\big)\big\{8\mathbb{C}_3\mathcal{C}^2
+2(\varpi+1)r^2g(g\beta(\mathcal{B}^2-2\mathcal{C}^2)+6)-g(\varpi+1)\\\nonumber
&\times(\mathcal{B}^2+2\mathcal{C}^2)\ln\big(\mathcal{B}^2+2r^2\big)(\mathcal{B}^2g\beta+4)+2
(\varpi+1)\beta{r^4g^2}\big\}+\frac{2\pi\varpi}{r^2}\\\nonumber
&\times\big(\mathcal{B}^2+2\mathcal{C}^2\big)-\frac{2(\mathcal{B}^2+2\mathcal{C}^2)}{(\mathcal{B}^2+2r^2)}
\big\{6\pi\varpi\big((\varpi+1)\mathcal{B}^2g-1\big)+(\varpi+1)\\\nonumber
&\times\varpi^2\mathcal{B}^2g+8\pi^2\big((\varpi+1)\mathcal{B}^2g-2\big)\big\}-\frac{1}{r^3}
\bigg\{\pi\varpi\sqrt{2\mathcal{B}^2}\big(\mathcal{B}^2+2\mathcal{C}^2\big)\\\label{60f}
&\times\tan^{-1}\bigg(r\sqrt{\frac{2}{\mathcal{B}^2}}\bigg)\bigg\}\bigg],
\end{align}
where $\mathbb{C}_3$ refers to an integration constant with
dimension $\frac{1}{l^2}$. Henceforth, the corresponding deformed
radial metric coefficient can be obtained as
\begin{align}\label{g60fa}
e^{\sigma}&=\rho^{-1}=\frac{\mathcal{C}^2\big(\mathcal{B}^2+2r^2\big)}{\big(\mathcal{B}^2+r^2\big)\big(\mathcal{C}^2-r^2\big)
+\beta\mathcal{C}^2\mathcal{F}(r)\big(\mathcal{B}^2+2r^2\big)}.
\end{align}
Finally, the second solution for the constraint
$\breve{\mathbb{Y}}_{TF}=0$ is given as
\begin{align}\nonumber
\breve{\mu}&=\frac{1}{64\mathcal{C}^2\big(\varpi+4\pi\big)\big(\varpi+2\pi\big)}
\bigg[\frac{16}{(\mathcal{B}^2+2r^2)^2}\big\{4 (\varpi +3 \pi )
\mathcal{B}^4+\mathcal{B}^2 \big(28 \pi r^2\\\nonumber &+8 \varpi
r^2+5 \varpi \mathcal{C}^2+12 \pi  \mathcal{C}^2\big)+2 r^2
\big(\varpi \big(3 r^2+2 \mathcal{C}^2\big)+4 \pi
\big(3r^2+\mathcal{C}^2\big)\big)\big\}\\\nonumber
&-\frac{\beta}{\pi  \big((\varpi +2) \mathcal{B}^2+(2 \varpi +3)
r^2\big)^2}\bigg\{2 r^2 \varrho_{1} \big((\varpi +2)
\mathcal{B}^2+(2 \varpi +3) r^2\big)\\\nonumber &+3 \varrho_{1}
\big(\mathcal{B}^2+r^2\big)\big((\varpi +2) \mathcal{B}^2+(2 \varpi
+3) r^2\big)-2 (2 \varpi +3) r^2 \varrho_{1}
\big(\mathcal{B}^2+r^2\big)\\\nonumber &+\frac{1}{r^3
\big(\mathcal{B}^2+r^2\big) \big(\mathcal{B}^2+2 r^2\big)^2
\big((\varpi +2) \mathcal{B}^2+(2 \varpi +3) r^2\big)}\bigg(8
r^5(\mathcal{B}^2+2 \mathcal{C}^2)\\\nonumber &\times  \big((\varpi
+1) \varpi ^2 \mathcal{B}^2 g+6 \pi \varpi ((\varpi +1)
\mathcal{B}^2g-1)+8 \pi ^2 ((\varpi +1)
\mathcal{B}^2g-2)\big)\\\nonumber &-4 \pi \varpi r
\big(\mathcal{B}^2+2 r^2\big)^2 (\mathcal{B}^2+2 \mathcal{C}^2)-2
\pi \varpi \mathcal{B}^2 r \big(\mathcal{B}^2+2 r^2\big)
(\mathcal{B}^2+2 \mathcal{C}^2)\\\nonumber &+3 \sqrt{2\mathcal{B}^2}
\pi \varpi \big(\mathcal{B}^2+2 r^2\big)^2 (\mathcal{B}^2+2
\mathcal{C}^2)\tan^{-1}\bigg(r\sqrt{\frac{2}{\mathcal{B}^2}}\bigg)+8
r^5g(\varpi +1)\\\nonumber &\times (\varpi +2 \pi ) (\varpi +4 \pi )
\big(\mathcal{B}^2+2 r^2\big) \big(\mathcal{B}^2 \big(2 \beta r^2
g-2 \beta g\mathcal{C}^2+1\big)+2 \beta  r^4 g\\\label{60g} &+r^2
(6-2\beta{g}\mathcal{C}^2)-4\mathcal{C}^2\big)\bigg)\bigg\}\bigg],\\\nonumber
\breve{P}_{r}&=\frac{1}{64\mathcal{C}^2\big(\varpi+2\pi\big)\big(\varpi+4\pi\big)}\bigg[\frac{16}{(\mathcal{B}^2+2r^2)^2}
\big\{\mathcal{B}^2 \big(4 \pi \big(\mathcal{C}^2-5 r^2\big)-4
\varpi  r^2\\\nonumber &+3 \varpi  \mathcal{C}^2\big)-4 \pi
\mathcal{B}^4+2 r^2 \big(-3 \varpi  r^2+4 \pi \big(\mathcal{C}^2-3
r^2\big)+2 \varpi \mathcal{C}^2\big)\big\}\\\nonumber
&+\frac{1}{\pi\beta\big(\varpi+2\pi\big)\big(\varpi+4\pi\big)\big(\mathcal{B}^2+r^2\big)}
\bigg\{\frac{16g\big(\mathcal{C}^2-r^2\big)}
{\mathcal{B}^2+2r^2}+\varrho_{1}\bigg(\frac{\mathcal{B}^2+3r^2}{\mathcal{B}^2+r^2}\bigg)\\\label{60h}
&\times\frac{1}{\big(\varpi+4\pi\big)\big(\varpi+2\pi\big)
\big((\varpi+2)\mathcal{B}^2+(2\varpi+3)r^2\big)}\bigg\}\bigg],\\\nonumber
\breve{P}_{\bot}&=\frac{1}{128\mathcal{C}^2\big(\mathcal{B}^2+2r^2\big)^2\big(\varpi+2\pi\big)\big(\varpi+4\pi\big)}
\bigg[32\big\{\mathcal{B}^2 \big(4 \pi \big(\mathcal{C}^2-5
r^2\big)\\\nonumber &-4 \varpi r^2+3 \varpi
\mathcal{C}^2\big)-4\pi\mathcal{B}^4+2 r^2 \big(4 \pi
\big(\mathcal{C}^2-3 r^2\big)-3 \varpi r^2+2 \varpi
\mathcal{C}^2\big)\big\}\\\nonumber &+\frac{1}{\pi  \beta
\big(\mathcal{B}^2+r^2\big) \big((\varpi +2) \mathcal{B}^2+(2 \varpi
+3) r^2\big)^2}\bigg\{16 (\varpi +2 \pi ) (\varpi +4 \pi )
\\\nonumber &\times g\big(\mathcal{B}^2+r^2\big) \big((\varpi +2)
\mathcal{B}^2+(2 \varpi +3) r^2\big)^2 \big(\mathcal{B}^4
\big(-gr^4+r^2 (g\mathcal{C}^2-3)\\\nonumber &+2
\mathcal{C}^2\big)+\mathcal{B}^2\big(-3 r^6 g+r^4 (3 g\mathcal{C}^2
-10)+7 r^2 \mathcal{C}^2\big)+2 r^6 (g\mathcal{C}^2-5)\\\nonumber
&-2 r^8 g+8 r^4 \mathcal{C}^2\big)+2 r^2 \varrho_{1} \big(2
\mathcal{B}^2+r^2\big) \big(\mathcal{B}^2+2
r^2\big)^2+r^{-3}\big(\mathcal{B}^2+r^2\big)\\\nonumber &\times
\big(\mathcal{B}^2+2 r^2\big) \big((\varpi +2) \mathcal{B}^2+(2
\varpi +3) r^2\big)\bigg(8 r^5(\mathcal{B}^2+2
\mathcal{C}^2)\big(\varpi ^2 \mathcal{B}^2\\\nonumber &\times
g(\varpi +1)+6 \pi \varpi ((\varpi +1) \mathcal{B}^2g-1)+8 \pi ^2
((\varpi +1) \mathcal{B}^2g-2)\big)\\\nonumber &-4 \pi \varpi r
\big(\mathcal{B}^2+2 r^2\big)^2 (\mathcal{B}^2+2 \mathcal{C}^2)-2
\pi \varpi \mathcal{B}^2 r \big(\mathcal{B}^2+2 r^2\big)
(\mathcal{B}^2+2 \mathcal{C}^2)\\\nonumber &+3 \sqrt{2\mathcal{B}^2}
\pi \varpi \big(\mathcal{B}^2+2 r^2\big)^2 (\mathcal{B}^2+2
\mathcal{C}^2)\tan^{-1}\bigg(r\sqrt{\frac{2}{\mathcal{B}^2}}\bigg)+8
r^5g(\varpi +1)\\\nonumber &\times (\varpi +2 \pi ) (\varpi +4 \pi )
\big(\mathcal{B}^2+2 r^2\big) \big(\mathcal{B}^2 \big(2 \beta r^2
g-2 \beta g\mathcal{C}^2+1\big)+2 \beta  r^4 g\\\nonumber &+r^2 (6-2
\beta{g}\mathcal{C}^2)-4\mathcal{C}^2\big)\bigg)-\big(\mathcal{B}^2+2
r^2\big)^3\big(2 (2\varpi +3) r^2
\varrho_{1}\big(\mathcal{B}^2+r^2\big)\\\nonumber &-2 r^2
\varrho_{1} \big((\varpi +2) \mathcal{B}^2+(2 \varpi +3) r^2\big)-2
\varrho_{1} \big(\mathcal{B}^2+r^2\big)\big((\varpi +2)
\mathcal{B}^2\\\label{60i} &+(2 \varpi +3)
r^2\big)\big)\bigg\}\bigg],
\end{align}
where $\varrho_{1}$ and the anisotropic factor for this solution are
given in Appendix \textbf{A}.

\section{Graphical Interpretation of the Obtained Solutions}

The following differential equation interlinks mass and energy
density of the sphere that can be solved numerically along with the
condition $m(0)=0$ to calculate mass as
\begin{equation}\label{g63}
\frac{dm(r)}{dr}=4\pi r^2 \breve{\mu},
\end{equation}
where $\breve{\mu}$ is provided in Eqs.\eqref{g46} and \eqref{60g}
with respect to each solution. The compactness of a geometrical
structure primarily measures how intricate particles of that object
are arranged, and we represent it as $\nu(r)$ in this case. The
tightness of these particles can be determined by calculating
compactness factor, which is also defined by the mass-radius ratio.
It was observed to be less than $\frac{4}{9}$ by Buchdahl \cite{42a}
for a spherical model. The noteworthy fact is that the wavelength of
electromagnetic radiations in the neighborhood of a self-gravitating
system is affected by its compactness. These radiations change their
path from geodesic motion due to the gravitational attraction of the
compact object. Such radiations are then redshifted which can
mathematically be determined as
\begin{equation}
z(r)=\big\{1-2\nu(r)\big\}^{-\frac{1}{2}}-1,
\end{equation}
whose maximum value is observed as 2 and 5.211 for perfect
\cite{42a} and anisotropic fluids \cite{42b}, respectively.

In the field of astrophysics, it is interesting to know whether a
compact structure incorporates an ordinary matter or not. The
presence of such a fluid and viability of the resulting model can be
guaranteed by the fulfillment of energy conditions. On the other
hand, if any of these bounds does not satisfy then there must be an
exotic matter inside the system. They have the following form in the
considered setup
\begin{eqnarray}\nonumber
&&\breve{\mu} \geq 0, \quad \breve{\mu}+\breve{P}_{r} \geq
0,\\\nonumber &&\breve{\mu}+\breve{P}_{\bot} \geq 0, \quad
\breve{\mu}-\breve{P}_{r} \geq 0,\\\label{g50}
&&\breve{\mu}-\breve{P}_{\bot} \geq 0, \quad
\breve{\mu}+\breve{P}_{r}+2\breve{P}_{\bot} \geq 0.
\end{eqnarray}
The obtained isotropic/anisotropic models which show stable behavior
are considered as an important subject of discussion among all the
celestial structures. Multiple techniques have been introduced in
this regard to check the stability. Firstly, we use the causality
condition according to which the inequalities $0<v_{s\bot}^{2}<1$
and $0<v_{sr}^{2}<1$ should be satisfied to get stable behavior
\cite{42bb}. Here,
$$v_{s\bot}^{2}=\frac{d\breve{P}_{\bot}}{d\breve{\mu}}, \quad v_{sr}^{2}=\frac{d\breve{P}_{r}}{d\breve{\mu}},$$ are the
tangential and radial sound speeds, respectively. If the absolute
value of the difference between both the sound speeds, i.e.,
$|v_{s\bot}^{2}-v_{sr}^{2}|$ lies in the interval $(0,1)$, then the
solution must be stable, named as the Herrera cracking concept
\cite{42ba}. Lastly, the adiabatic index ($\Gamma$) is analyzed
which states that the stable star must satisfy the inequality
$\Gamma>\frac{4}{3}$ \cite{42c}. Here, $\breve{\Gamma}$ in terms of
effective energy density and pressure is given as
\begin{equation}\label{g62}
\breve{\Gamma}=\frac{\breve{\mu}+\breve{P}_{r}}{\breve{P}_{r}}
\bigg(\frac{d\breve{P}_{r}}{d\breve{\mu}}\bigg).
\end{equation}

We take a linear $f(\mathbb{R},\mathbb{T})$ model \eqref{g61} to
explore deformation functions, the scalar $\mathbb{Y}_{TF}$ and the
effective state variables corresponding to different constraints
through graphical analysis. We consider multiple choices of the
decoupling and coupling parameters to study their role on physical
attributes of a particular $4U 1820-30$ model by fixing
$\mathbb{C}_{1}=0.01$ and $g=0.003$. Figure \textbf{1} manifests
graphs of the deformed $g_{rr}$ metric function \eqref{g40aa}
showing increasing and non-singular trend for $0<r<\mathcal{R}$. The
parameters governing the matter distribution (such as pressure and
energy density) are required to be maximum, positive and finite at
the center, and show linear decrement towards the boundary to meet
acceptability criteria of the developed model. We observe acceptable
behavior of our first solution \eqref{g46}-\eqref{g49} from Figure
\textbf{2}. The upper left plot shows that the effective energy
density is maximum in the core. Further, it decreases by increasing
both the parameters $\varpi$ and $\beta$ near the center, while
shows opposite trend near the boundary. On the other hand, the
pressure in both directions is observed to be in direct relation
with $\varpi$ as well as $\beta$. The vanishing of the radial
pressure at $r=\mathcal{R}$ can also be seen from the upper right
plot. We transform anisotropic system to an isotropic for $\beta=1$
that can be confirmed  from the last plot, as the anisotropy is zero
throughout for that value. Moreover, this factor shows increasing
behavior for the remaining values of $\beta$.
\begin{figure}\center
\epsfig{file=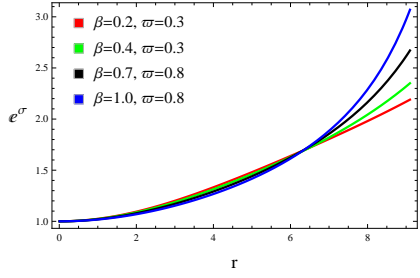,width=0.4\linewidth}
\caption{Deformed $g_{rr}$ component \eqref{g40aa} for the solution
corresponding to $\breve{\Pi}=0$.}
\end{figure}
\begin{figure}\center
\epsfig{file=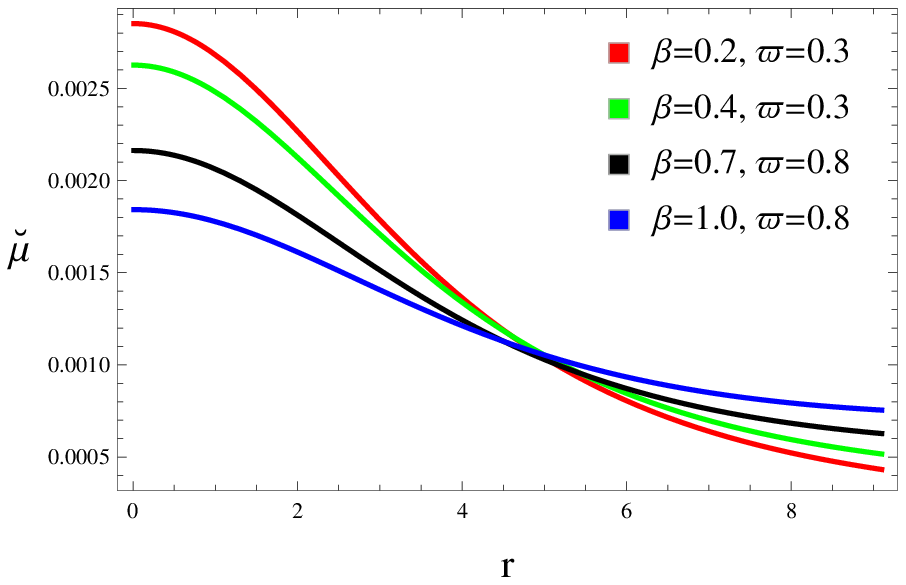,width=0.4\linewidth}\epsfig{file=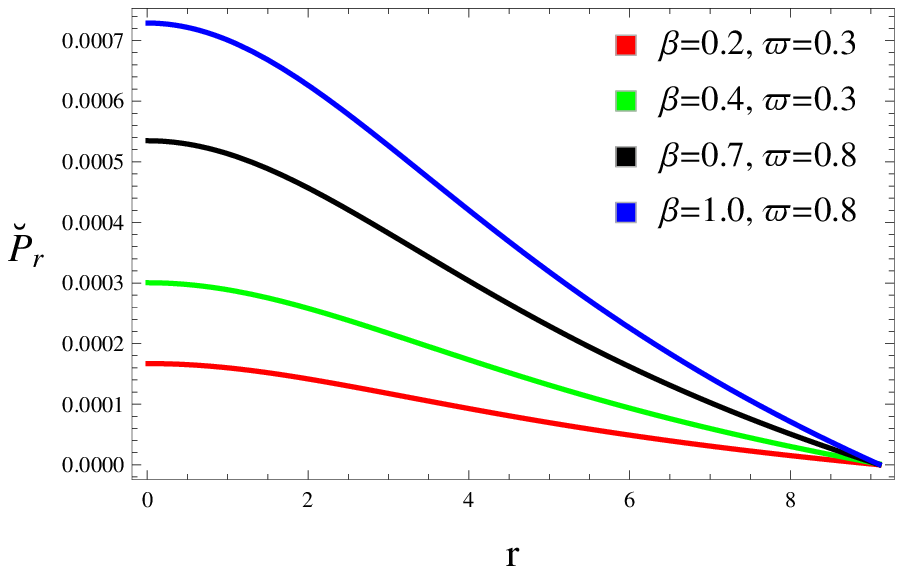,width=0.4\linewidth}
\epsfig{file=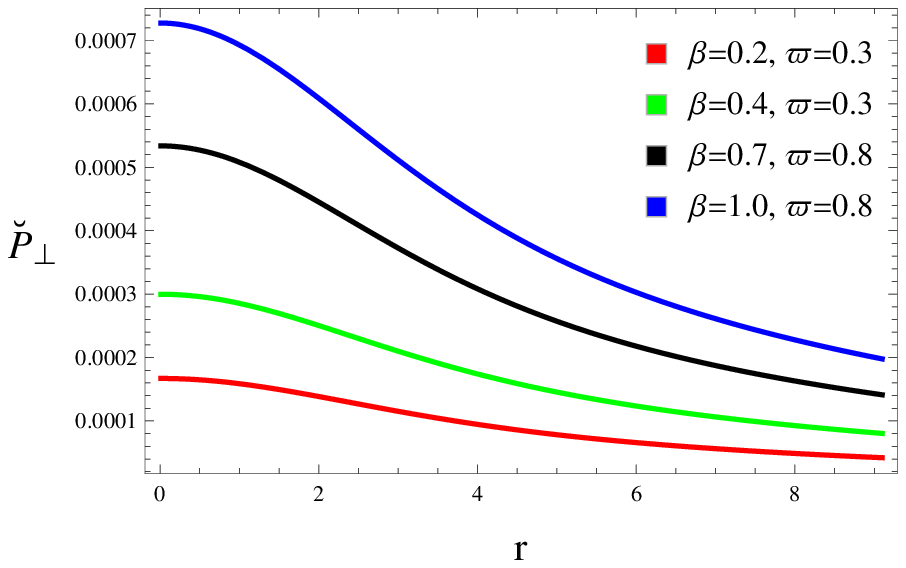,width=0.4\linewidth}\epsfig{file=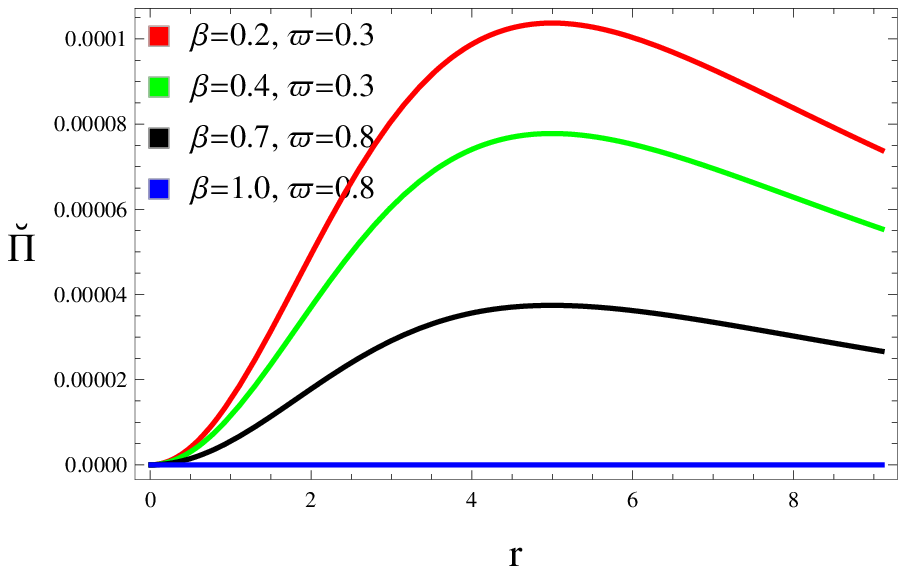,width=0.4\linewidth}
\caption{Matter variables and pressure anisotropy for the solution
corresponding to $\breve{\Pi}=0$.}
\end{figure}
\begin{figure}\center
\epsfig{file=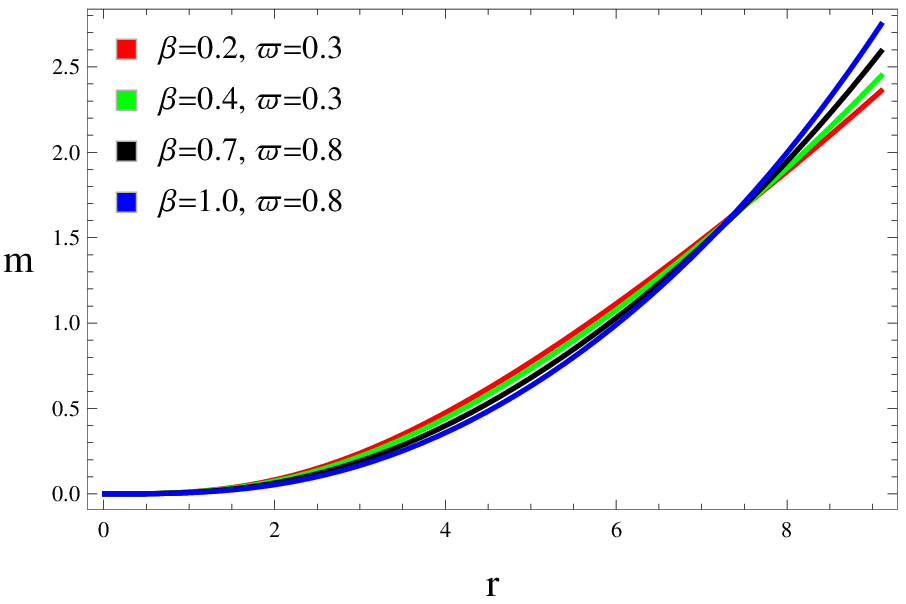,width=0.4\linewidth}\epsfig{file=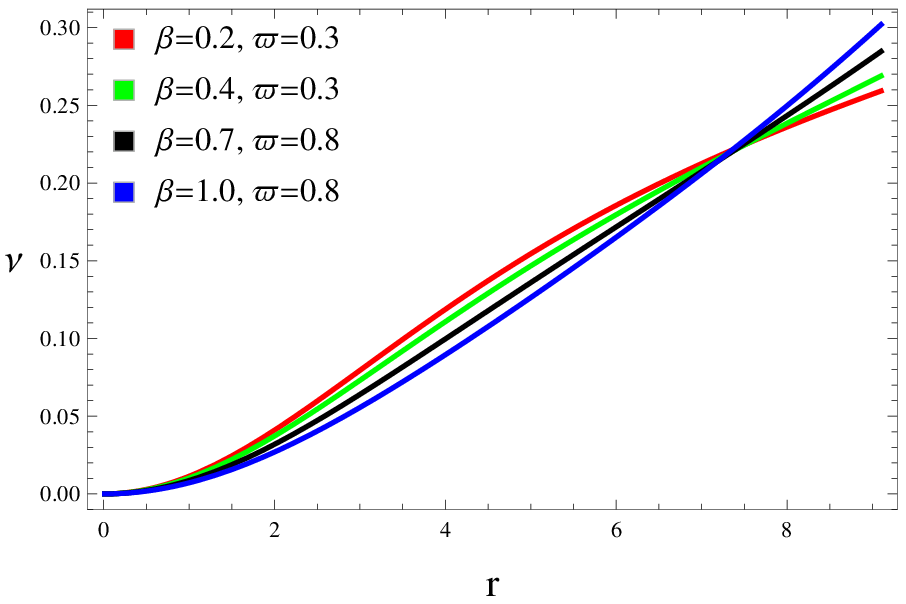,width=0.4\linewidth}
\epsfig{file=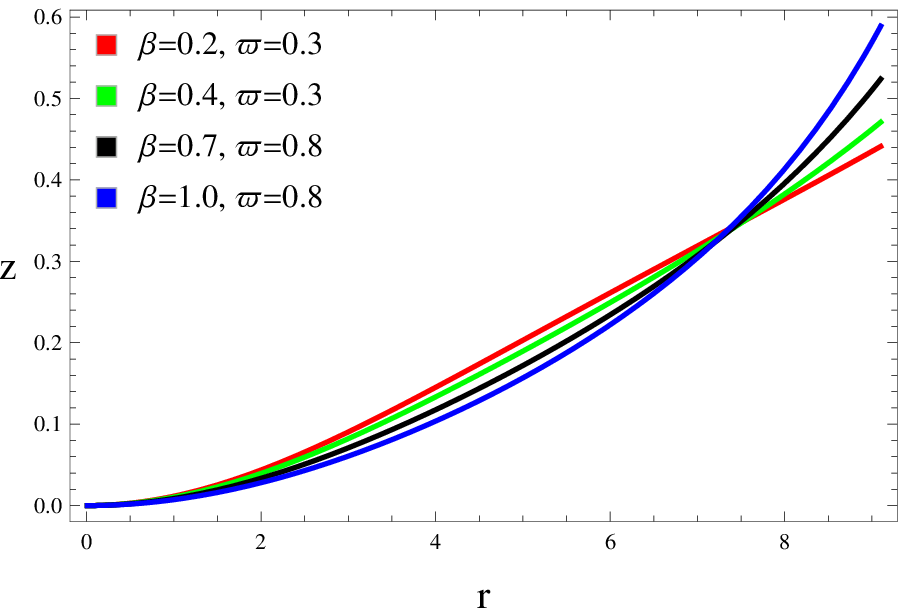,width=0.4\linewidth} \caption{Mass,
compactness and redshift for the solution corresponding to
$\breve{\Pi}=0$.}
\end{figure}
\begin{figure}\center
\epsfig{file=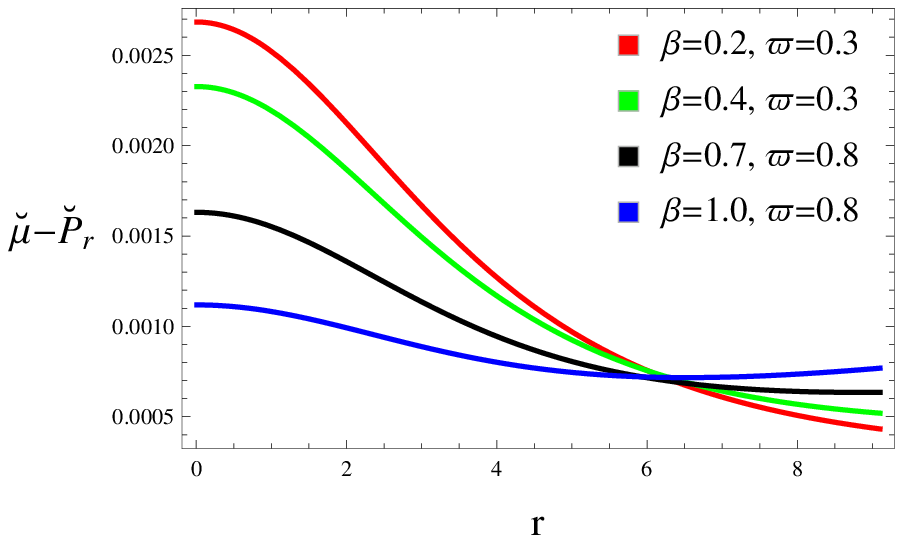,width=0.4\linewidth}\epsfig{file=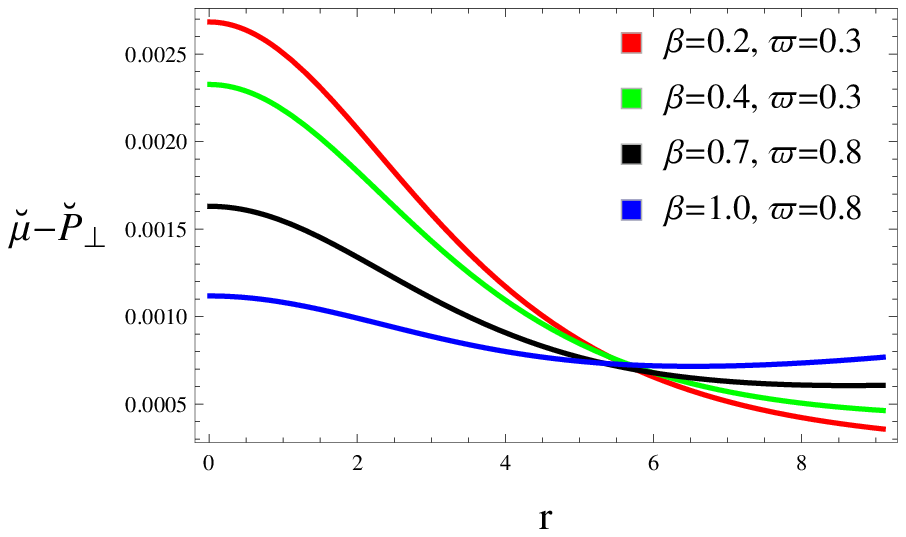,width=0.4\linewidth}
\caption{Dominant energy conditions for the solution corresponding
to $\breve{\Pi}=0$.}
\end{figure}
\begin{figure}\center
\epsfig{file=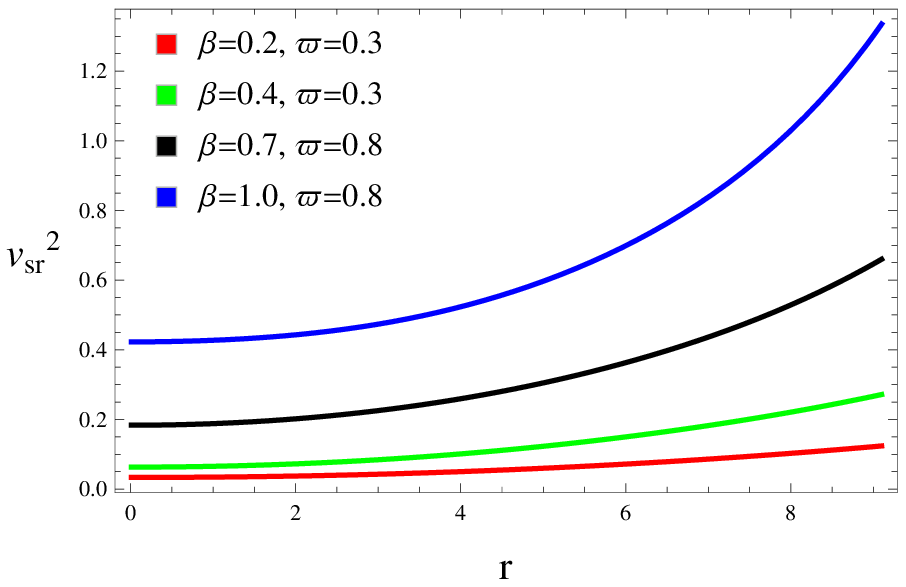,width=0.4\linewidth}\epsfig{file=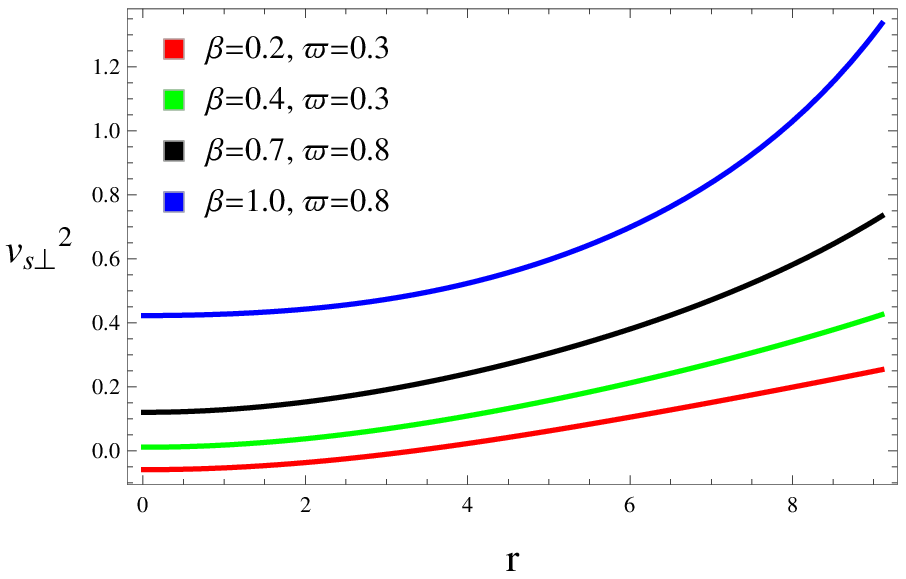,width=0.4\linewidth}
\epsfig{file=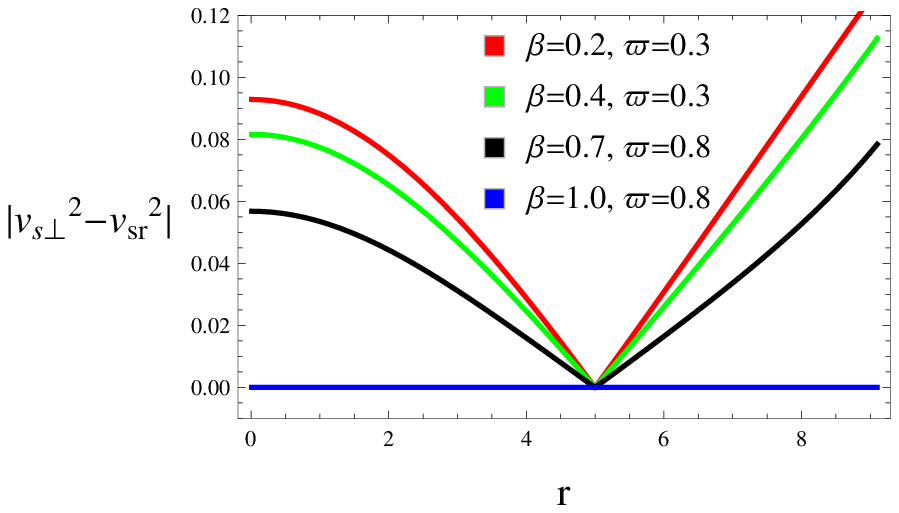,width=0.4\linewidth}\epsfig{file=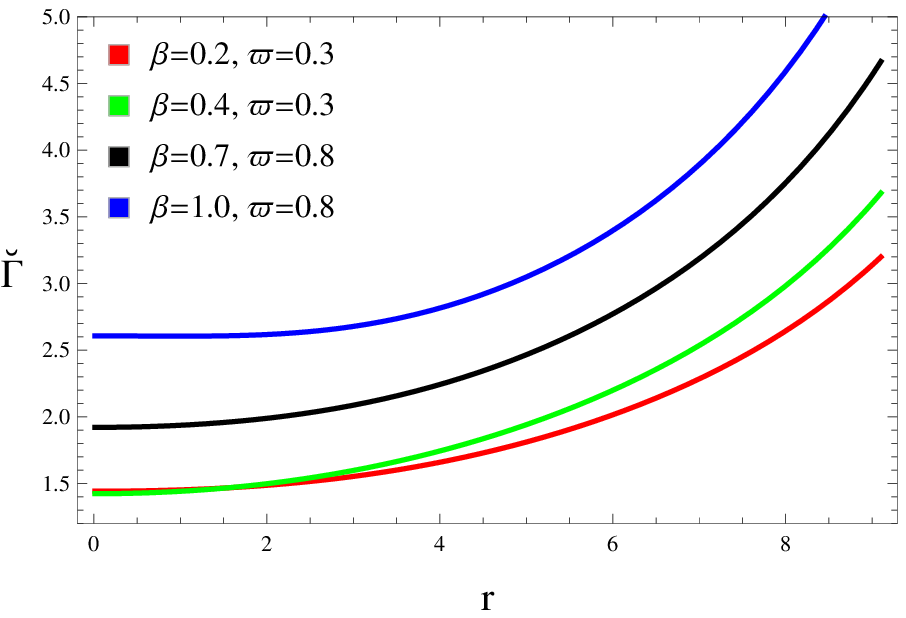,width=0.4\linewidth}
\caption{Radial/tangential velocities,~$|v_{s\bot}^2-v_{sr}^2|$ and
adiabatic index for the solution corresponding to $\breve{\Pi}=0$.}
\end{figure}

Figure \textbf{3} (upper left plot) reveals that the isotropic
system is initially less massive than anisotropic analog, but
exhibits counter consequence towards the hypersurface. The
compactness and redshift are plotted in the upper right and lower
plots, respectively which meet their required limits. All the energy
bounds must be fulfilled due to positive nature of the state
variables except dominant conditions ($\breve{\mu}-\breve{P}_{r}
\geq 0$ and $\breve{\mu}-\breve{P}_{\bot} \geq 0$), thus they are
needed to be checked. Figure \textbf{4} assures validity of these
conditions, hence our first solution is physically viabile. We check
the stability by using three different approaches in Figure
\textbf{5}. The developed model \eqref{g46}-\eqref{g49} is unstable
near the center for $\beta=0.2$ (upper left plot) as well as near
the boundary for $\beta=1$ (upper two plots), while becomes stable
for all the remaining considered choices of $\varpi$ and $\beta$ .
Meanwhile, the resulting solution is observed to be stable
everywhere by using the adiabatic index and the cracking condition
(lower two plots). Figure \textbf{6} indicates that the increment in
both $\varpi$ and $\beta$ decreases the complexity factors
\eqref{g56a} and \eqref{g60d} which follows that the impact of
complexity in $f(\mathbb{R},\mathbb{T})$ theory is much lesser than
that of $\mathbb{GR}$.
\begin{figure}\center
\epsfig{file=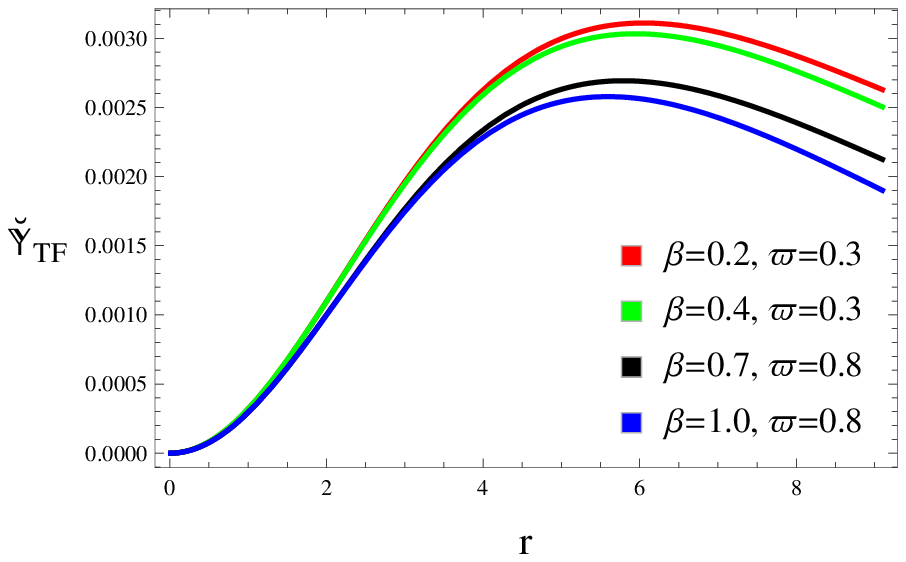,width=0.4\linewidth}\epsfig{file=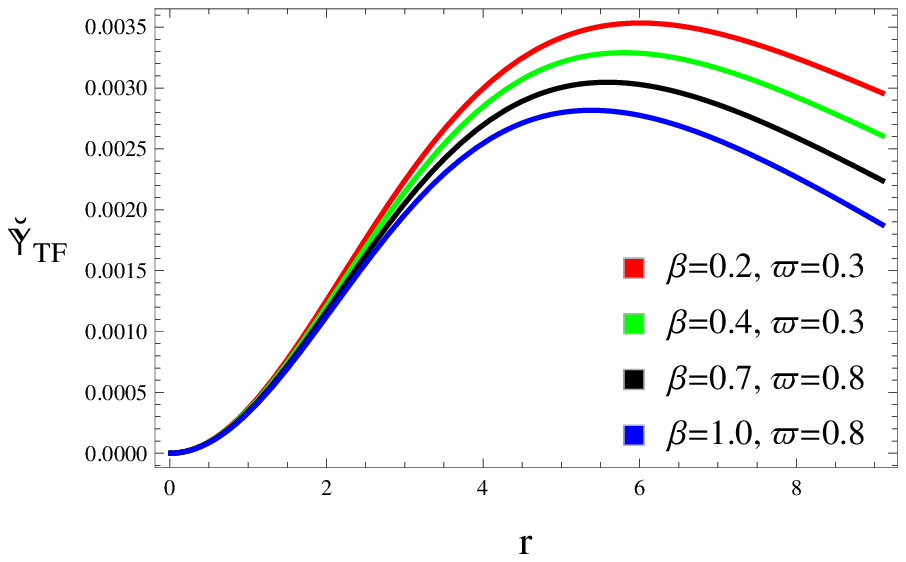,width=0.4\linewidth}
\caption{Complexity factors \eqref{g56a} and \eqref{g60d}.}
\end{figure}
\begin{figure}\center
\epsfig{file=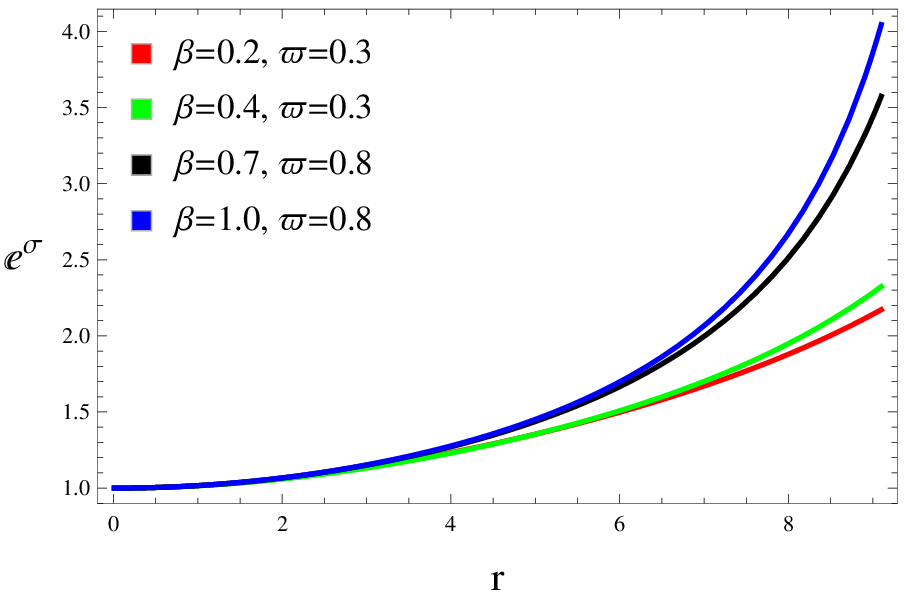,width=0.4\linewidth}
\caption{Deformed $g_{rr}$ component \eqref{g60fa} for the solution
corresponding to $\breve{\mathbb{Y}}_{TF}=0$.}
\end{figure}

We adopt another constraint $\breve{\mathbb{Y}}_{TF}=0$ and obtain
the corresponding solution \eqref{60g}-\eqref{60i} whose physical
characteristics are analyzed by choosing $\mathbb{C}_{3}=0.01$.
Figure \textbf{7} provides the plot of deformed $g_{rr}$ component
presenting non-singular trend throughout. The matter variables and
pressure anisotropy are analyzed in Figure \textbf{8} which show
acceptable behavior as we have discussed earlier. The tangential
pressure in this case is lesser than the radial component (except at
the center), thus anisotropy appears to be negative throughout.
Figure \textbf{9} (upper left) exhibits that the sphere becomes more
massive by increasing both parameters $\varpi$ and $\beta$, in
contrast to the result obtained through MGD. The right and lower
plots confirm the acceptance of other two factors. Figure
\textbf{10} discloses that our second solution as well as extended
model \eqref{g61} are also viable. Figure \textbf{11} reveals the
stability of this solution everywhere.
\begin{figure}\center
\epsfig{file=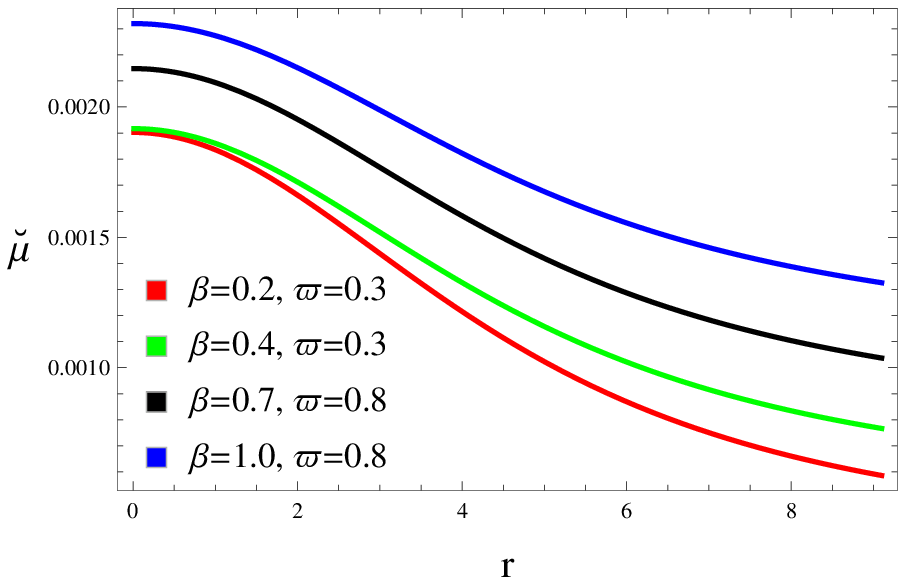,width=0.4\linewidth}\epsfig{file=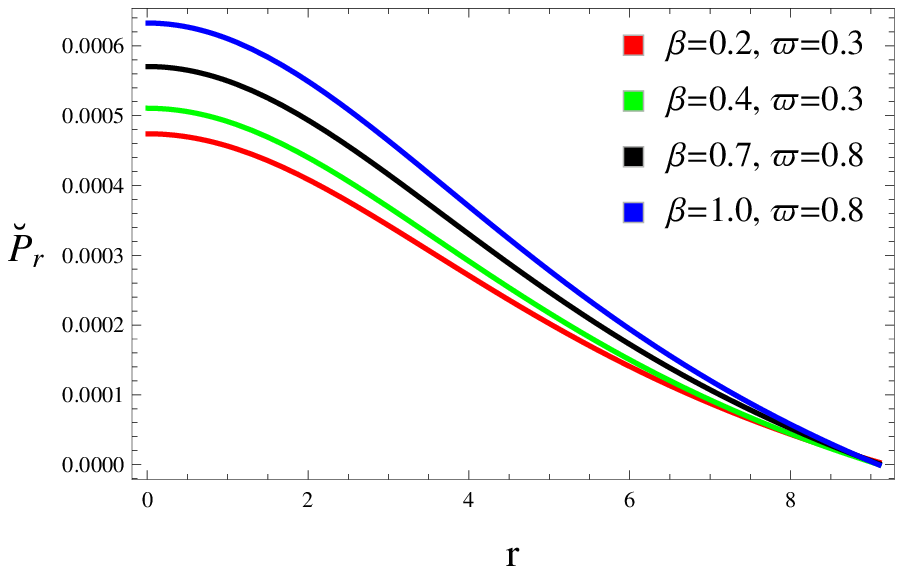,width=0.4\linewidth}
\epsfig{file=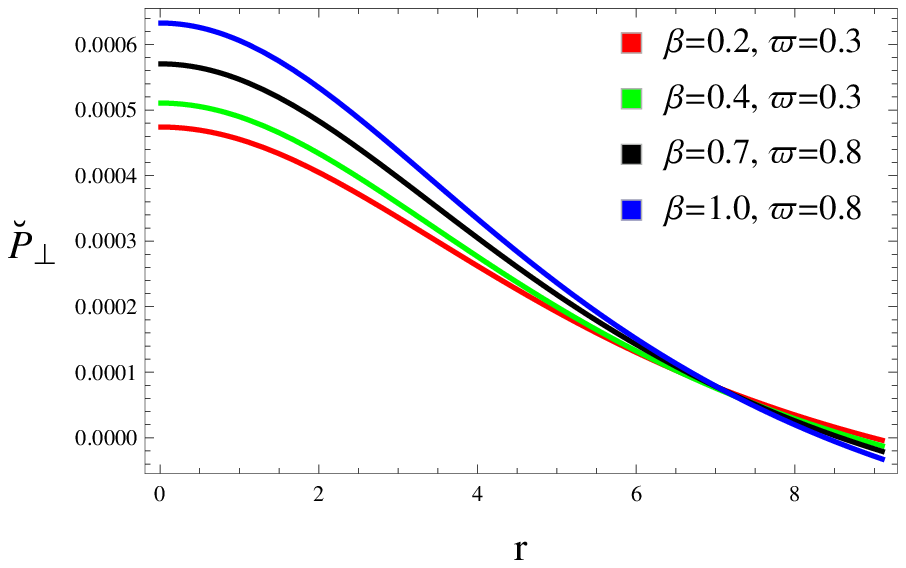,width=0.4\linewidth}\epsfig{file=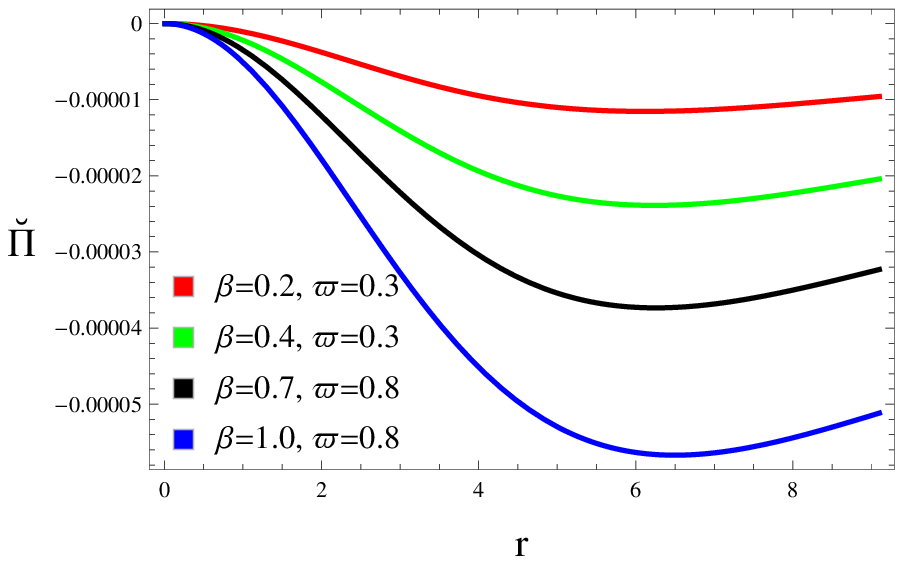,width=0.4\linewidth}
\caption{Matter variables and pressure anisotropy for the solution
corresponding to $\breve{\mathbb{Y}}_{TF}=0$.}
\end{figure}
\begin{figure}\center
\epsfig{file=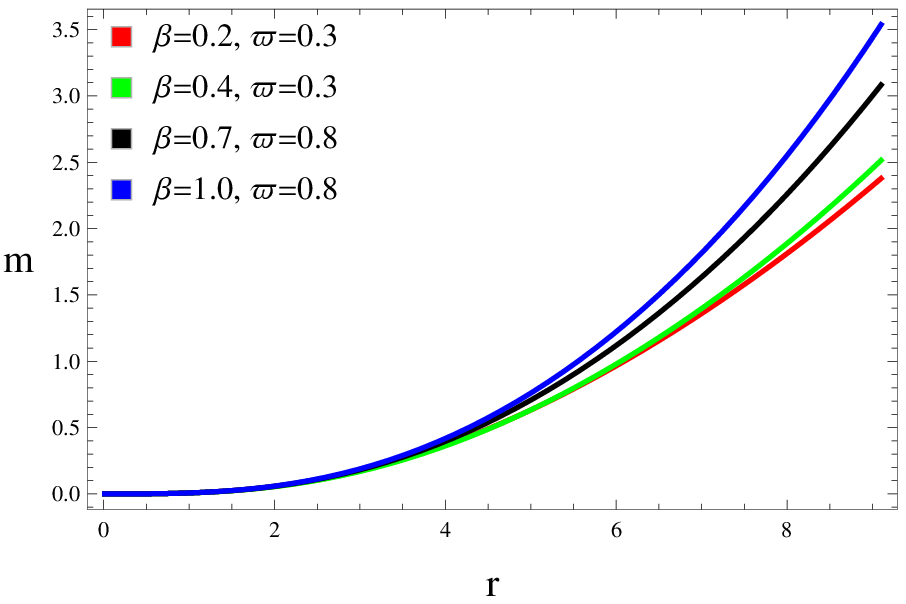,width=0.4\linewidth}\epsfig{file=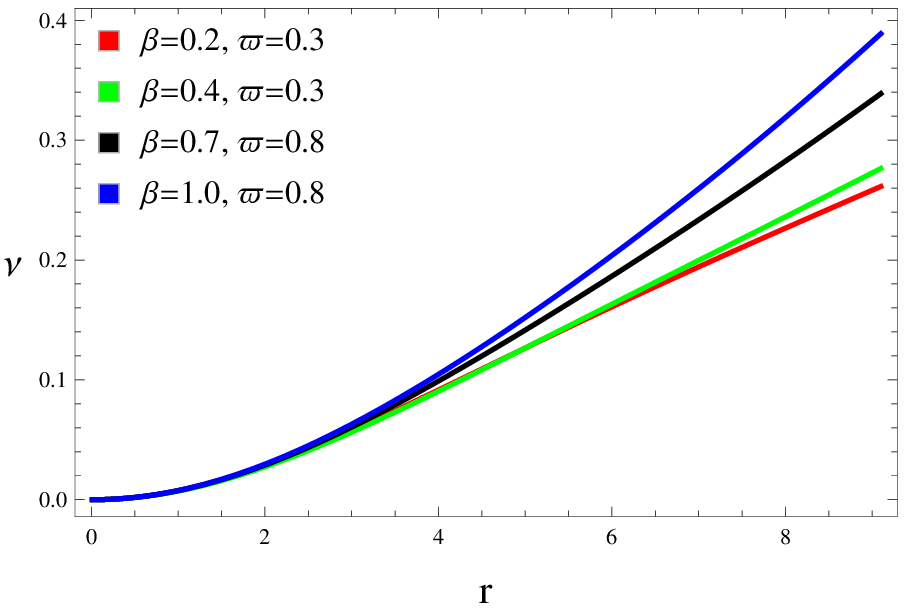,width=0.4\linewidth}
\epsfig{file=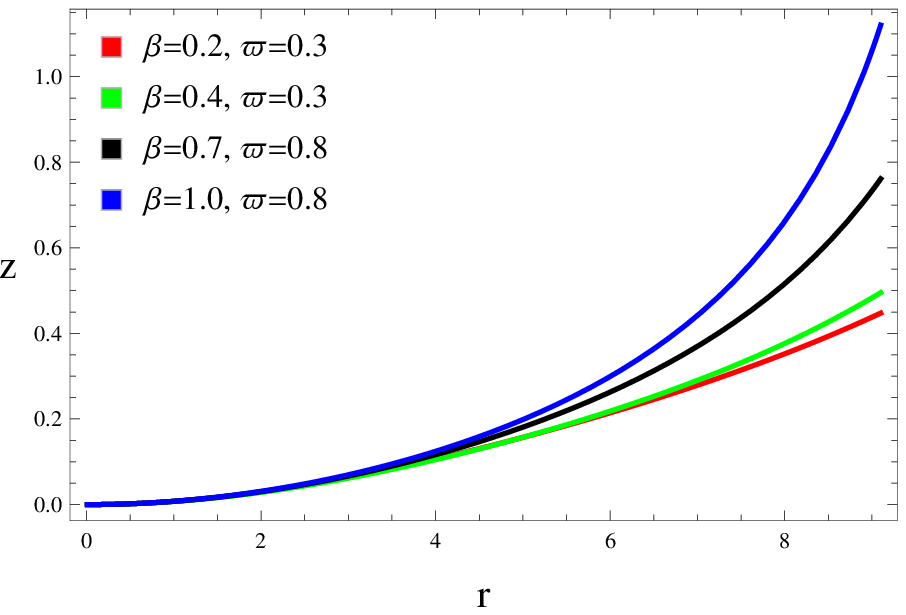,width=0.4\linewidth} \caption{Mass,
compactness and redshift for the solution corresponding to
$\breve{\mathbb{Y}}_{TF}=0$.}
\end{figure}
\begin{figure}\center
\epsfig{file=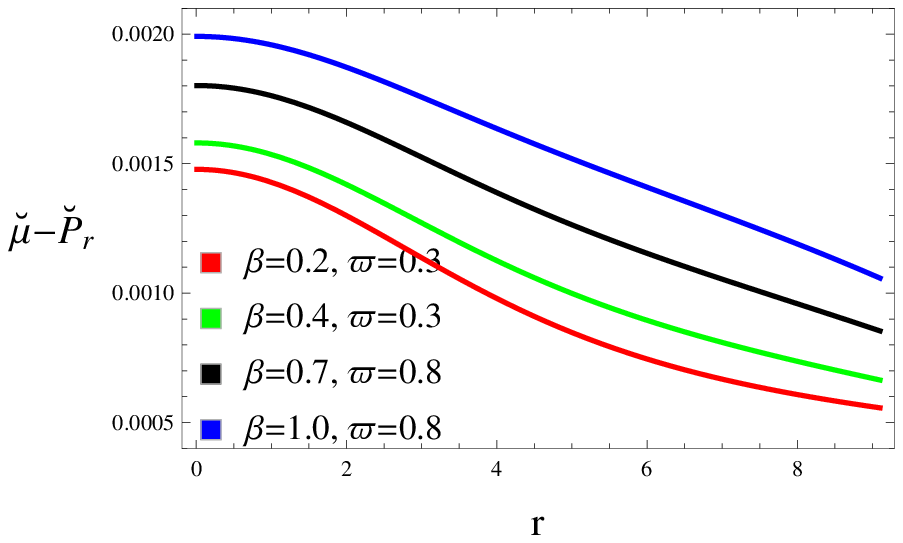,width=0.4\linewidth}\epsfig{file=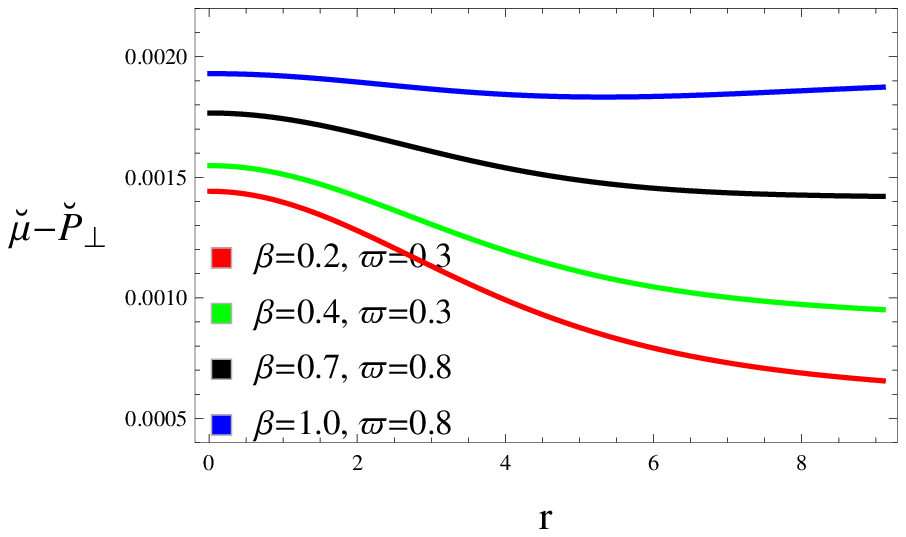,width=0.4\linewidth}
\caption{Dominant energy conditions for the solution corresponding
to $\breve{\mathbb{Y}}_{TF}=0$.}
\end{figure}
\begin{figure}\center
\epsfig{file=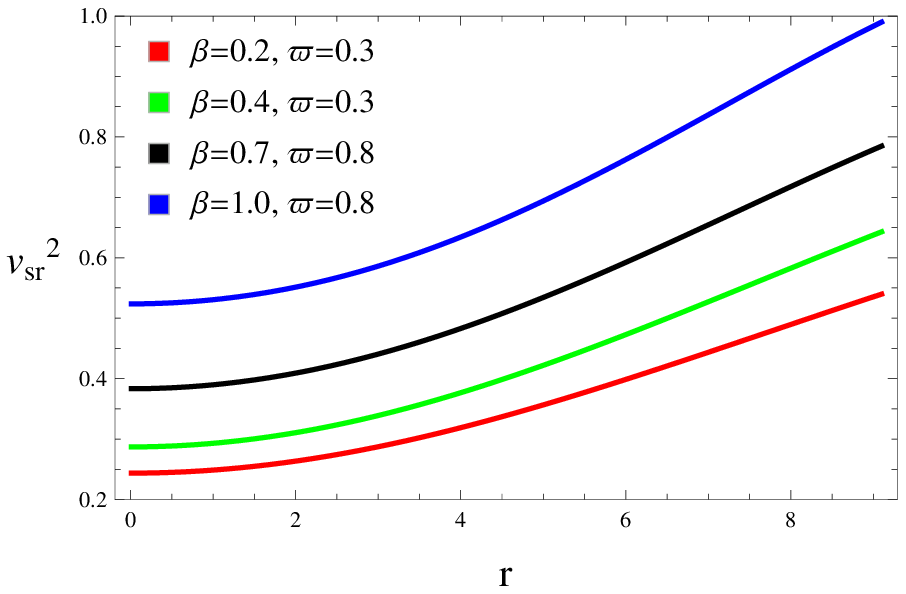,width=0.4\linewidth}\epsfig{file=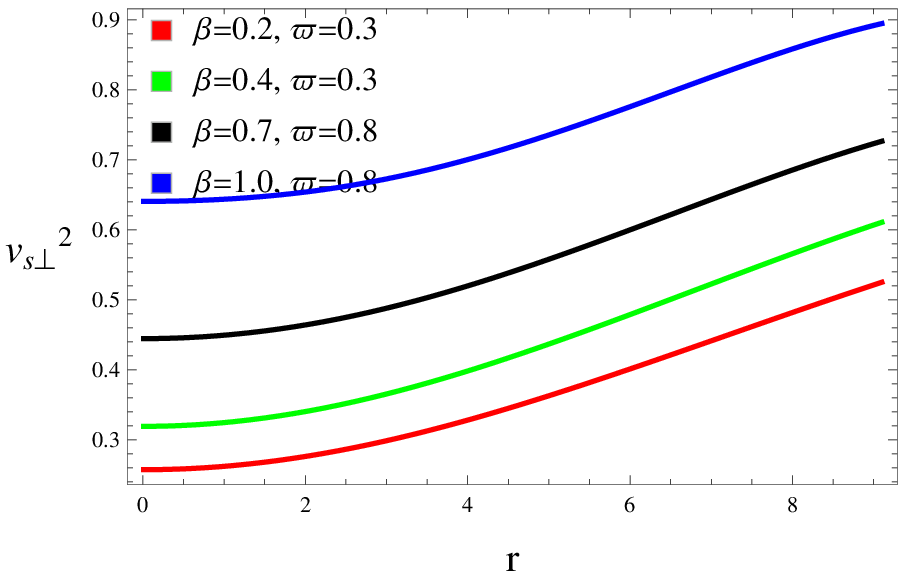,width=0.4\linewidth}
\epsfig{file=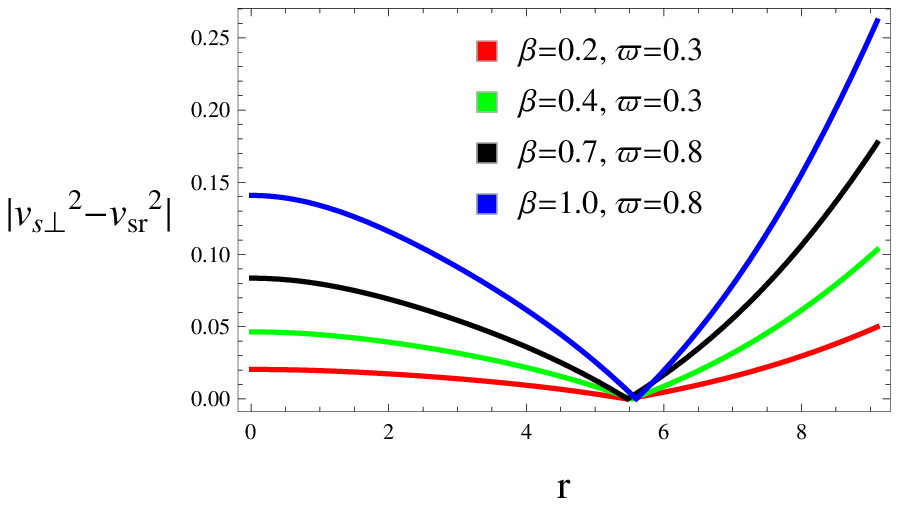,width=0.4\linewidth}\epsfig{file=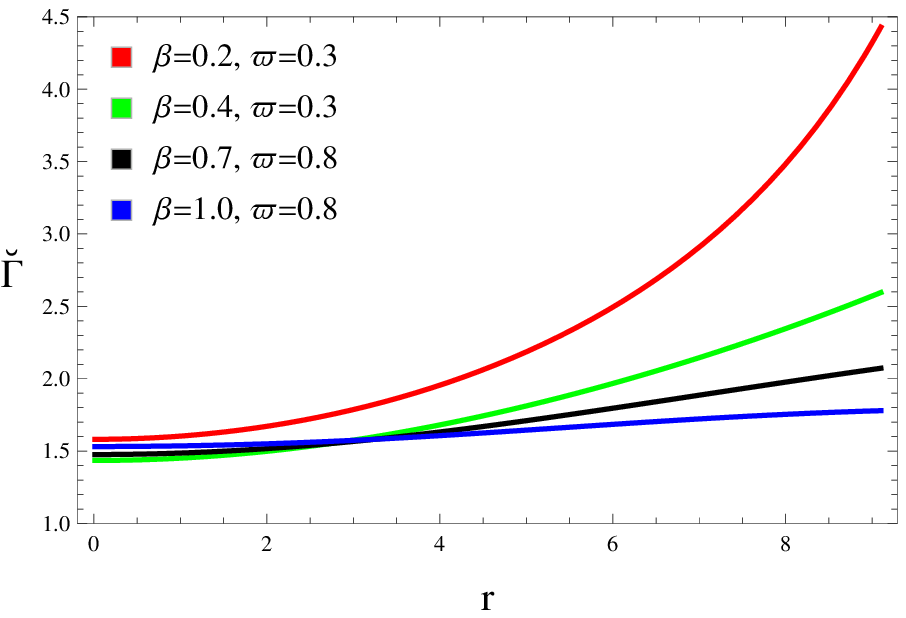,width=0.4\linewidth}
\caption{Radial/tangential velocities,~$|v_{s\bot}^2-v_{sr}^2|$ and
adiabatic index for the solution corresponding to
$\breve{\mathbb{Y}}_{TF}=0$.}
\end{figure}

\section{Conclusions}

The main objective of this paper is to extend some solutions for
anisotropic self-gravitating distribution ($\mathbb{T}_{\eta\zeta}$)
through the addition of an extra source ($\mathfrak{B}_{\eta\zeta}$)
by employing gravitational decoupling scheme in
$f(\mathbb{R},\mathbb{T})=\mathbb{R}+2\varpi\mathbb{T}$ gravity. The
corresponding field equations representing the total fluid
configuration (seed and additional sources) have been formulated
which are then split into two distinct sets by means of an EGD
technique. Both sets of equations have ultimately represented their
parent sources. For the first set describing seed source, we have
taken the following metric potentials
$$\chi(r)=\ln\bigg\{\mathcal{A}^2\bigg(1+\frac{r^2}{\mathcal{B}^2}\bigg)\bigg\},\quad
\rho(r)=e^{-\sigma(r)}=\frac{\mathcal{B}^2+r^2}{\mathcal{B}^2+3r^2},$$
and the Tolman IV ansatz, resulting in two extended solutions. We
have used the radius and mass of compact $4U 1820-30$ model in order
to determine three unknowns involving in these components. There
have been five unknowns
($\mathcal{F},\mathcal{G},\mathfrak{B}_{00},\mathfrak{B}_{11},\mathfrak{B}_{22}$)
incorporated in Eqs.\eqref{g21}-\eqref{g23}, thus we required to
implement two constraints simultaneously to get the solution. We
have considered a particular form of $\mathcal{G}(r)$ along with two
independent constraints leading to first and second solution. One
constraint has been taken as disappearance of the effective
anisotropy for $\beta=1$, therefore the system becomes an isotropic
for this value. We have then implemented another limitation on the
complexity, i.e., the total setup has been considered to be the
complexity-free.

The role of modified gravity and gravitational decoupling on
physical attributes of the developed models have been explored by
choosing $\varpi=0.3,~0.8$ and $\beta=0.2,~0.4,~0.7,~1$. The
graphical interpretation of corresponding state determinants
\big(\eqref{g46}-\eqref{g48} and \eqref{60g}-\eqref{60i}\big),
anisotropic pressure \big(\eqref{g49} and \eqref{A2}\big) and the
energy constraints \eqref{g50} have provided acceptable results for
specific values of multiple constants. The redshift and compactness
have also been shown to be within their required limits (Figures
\textbf{3} and \textbf{9}). We have observed that the obtained model
corresponding to $\breve{\mathbb{Y}}_{TF}=0$ yields denser stellar
structure, in comparison with the other solution. All the
deformation functions have been found to be zero in the core of
considered star and increasing towards the boundary. Further, the
solution corresponding to $\breve{\Pi}$ is stable throughout only
for $\beta=0.4,~0.7$ with causality condition, whereas the other
solution is stable everywhere (Figures \textbf{5} and \textbf{11}).
The adiabatic index and Herrera's cracking concept have provided
stable structures, thus our obtained solutions are compatible with
those of $\mathbb{GR}$ \cite{37k}. The constraint $\breve{\Pi}=0$
has also been used in the the Brans-Dicke scenario to get solution
which shows inconsistent behavior with $f(\mathbb{R},\mathbb{T})$
gravity, as $\beta=1$ provides unstable system in this case
\cite{37m}. Finally, $\varpi=0$ reduces all of our results to
$\mathbb{GR}$.

\section*{Appendix A}

\renewcommand{\theequation}{A\arabic{equation}}
\setcounter{equation}{0} The value of $\varrho_{1}$ appearing in
Eqs.\eqref{60g}-\eqref{60i} is
\begin{align}\nonumber
\varrho_{1}&=(\varpi +2 \pi ) (\varpi +4 \pi ) \big\{2 (\varpi +1)
r^2 g (\beta g (\mathcal{B}^2-2 \mathcal{C}^2)+6)-(\varpi +1) g
\\\nonumber &\times (\mathcal{B}^2+2 \mathcal{C}^2)\ln
\big(\mathcal{B}^2+2 r^2\big) (\mathcal{B}^2 \beta g+4)+8
\mathbb{C}_3 \mathcal{C}^2+2 (\varpi +1) \beta  r^4
g^2\big\}\\\nonumber &+\frac{2\pi\varpi(\mathcal{B}^2+2
\mathcal{C}^2)}{r^2}-\frac{1}{(\mathcal{B}^2+2r^2)}\big\{2
(\mathcal{B}^2+2 \mathcal{C}^2)\big((\varpi+1) \varpi^2
\mathcal{B}^2 g \\\nonumber &+6 \pi \varpi((\varpi +1) \mathcal{B}^2
g-1)+8 \pi ^2 ((\varpi +1) \mathcal{B}^2
g-2)\big)\big\}-\frac{\pi\varpi\sqrt{2\mathcal{B}^2}}{r^3}\\\label{A1}
&\times(\mathcal{B}^2+2 \mathcal{C}^2)\tan ^{-1}\bigg(\frac{\sqrt{2}
r}{\sqrt{\mathcal{B}^2}}\bigg).
\end{align}

\noindent The anisotropy for our second solution
\eqref{60g}-\eqref{60i} is
\begin{align}\nonumber
\breve{\Pi}&=\frac{\beta}{8\pi}\bigg[\frac{2g\big(\mathcal{B}^2+r^2\big)
\big(r^2-\mathcal{C}^2\big)}{\mathcal{C}^2\big(\mathcal{B}^2
+2r^2\big)}+\frac{g}{\mathcal{C}^2\big(\mathcal{B}^2+2r^2\big)^2}\big\{\mathcal{B}^4\big(r^2(g\mathcal{C}^2-3)\\\nonumber
&-r^4g+2\mathcal{C}^2\big)+\mathcal{B}^2\big(-3r^6g+r^4(3g\mathcal{C}^2-10)+7r^2\mathcal{C}^2\big)-2r^8g\\\nonumber
&+2r^6(g\mathcal{C}^2-5)+8r^4\mathcal{C}^2\big\}-\frac{2\mathcal{B}^2r^2+\mathcal{B}^4+3r^4}{8\big(\mathcal{B}^2+r^2\big)
\big\{(\varpi+2)\mathcal{B}^2+(2\varpi+3)r^2\big\}}\\\nonumber
&\times\bigg(\frac{\varrho_{2}}{\mathcal{C}^2\big(\varpi+2\pi\big)\big(\varpi+4\pi\big)}+8\mathbb{C}_3\bigg)
+\big\{16(\varpi+2\pi\big)\big(\varpi+4\pi\big)\\\nonumber
&\times\big(\mathcal{B}^2+r^2\big)\big((\varpi+2)\mathcal{B}^2+(2\varpi+3)r^2\big)\big\}^{-1}\big(\mathcal{B}^2+2r^2\big)
\bigg\{\mathcal{C}^2 \big(16 \mathbb{C}_3 r^2\\\nonumber &\times
\big((\varpi +2) \mathcal{B}^2+(2 \varpi +3) r^2\big)-16 (2 \varpi
+3) \mathbb{C}_3 r^2 \big(\mathcal{B}^2+r^2\big)+16
\mathbb{C}_3\\\nonumber &\times \big(\mathcal{B}^2+r^2\big)
\big((\varpi +2) \mathcal{B}^2+(2 \varpi +3)
r^2\big)\big)\big(\varpi ^2+6 \pi \varpi +8 \pi ^2\big)-2r^2
\\\nonumber &\times(2 \varpi +3) \varrho_{2} \big(\mathcal{B}^2+r^2\big)+2 r^2
\varrho_{2} \big((\varpi +2) \mathcal{B}^2+(2 \varpi +3) r^2\big)+2
\varrho_{2}\\\nonumber &\times \big(\mathcal{B}^2+r^2\big)
\big((\varpi+2)\mathcal{B}^2+(2\varpi+3)r^2\big)+r\big(\mathcal{B}^2+r^2\big)\big((\varpi+2)\mathcal{B}^2\\\nonumber
&+(2\varpi+3)r^2\big)\bigg(4 (\varpi +1) \big(\varpi ^2+6 \pi \varpi
+8 \pi ^2\big) (\mathcal{B}^2 \beta  g-2 \beta
g\mathcal{C}^2+6)\\\nonumber &\times r g +8\beta  r^3 g^2 (\varpi
+1) \big(\varpi ^2+6 \pi \varpi +8 \pi ^2\big)-\frac{4 \pi
\varpi(\mathcal{B}^2+2 \mathcal{C}^2)}{r^3} \\\nonumber &-\frac{2\pi
\varpi\mathcal{B}^2(\mathcal{B}^2+2 \mathcal{C}^2)}{r^3
\big(\mathcal{B}^2+2 r^2\big)}+3 r^{-4}\sqrt{2\mathcal{B}^2} \pi
\varpi (\mathcal{B}^2+2 \mathcal{C}^2) \tan
^{-1}\bigg(\frac{\sqrt{2} r}{\sqrt{\mathcal{B}^2}}\bigg)\\\nonumber
&+8 r (\mathcal{B}^2+2 \mathcal{C}^2)(\mathcal{B}^2+2r^2)^{-2}
\big(6 \pi \varpi \big((\varpi+1)\mathcal{B}^2g-1\big)+8 \pi ^2
((\varpi +1)\\\nonumber &\times \mathcal{B}^2g-2)+(\varpi+1)
\varpi^2 \mathcal{B}^2g\big)-4 (\varpi+1)(\mathcal{B}^2+2r^2)^{-1}
(\mathcal{B}^2+2 \mathcal{C}^2)\\\label{A2} &\times
rg\big(\varpi^2+6 \pi \varpi +8 \pi ^2\big) (\mathcal{B}^2\beta
g+4)\bigg)\bigg\}\bigg],
\end{align}
where
\begin{align}\nonumber
\varrho_{2}&=(\varpi +2 \pi ) (\varpi +4 \pi ) \big\{2 (\varpi +1)
r^2 g (\beta g (\mathcal{B}^2-2 \mathcal{C}^2)+6)+2\beta g^2
\\\nonumber &\times(\varpi +1)  r^4 -(\varpi +1) g (\mathcal{B}^2+2 \mathcal{C}^2)\ln
\big(\mathcal{B}^2+2 r^2\big) (\mathcal{B}^2 \beta
g+4)\big\}\\\nonumber &+\frac{2\pi\varpi(\mathcal{B}^2+2
\mathcal{C}^2)}{r^2}-\frac{1}{(\mathcal{B}^2+2r^2)}\big\{2
(\mathcal{B}^2+2 \mathcal{C}^2)\big((\varpi+1) \varpi^2
\mathcal{B}^2 g \\\nonumber &+6 \pi \varpi((\varpi +1) \mathcal{B}^2
g-1)+8 \pi ^2 ((\varpi +1) \mathcal{B}^2
g-2)\big)\big\}-\frac{\pi\varpi\sqrt{2\mathcal{B}^2}}{r^3}\\\label{A3}
&\times(\mathcal{B}^2+2 \mathcal{C}^2)\tan ^{-1}\bigg(\frac{\sqrt{2}
r}{\sqrt{\mathcal{B}^2}}\bigg).
\end{align}

\end{document}